\documentclass[a4paper,fleqn]{aa}
\pdfoutput=1
\usepackage{natbib}
\usepackage[varg]{txfonts}
\usepackage[hidelinks]{hyperref}
\usepackage[dvipsnames]{xcolor}
\hypersetup{
    colorlinks,
    linkcolor={blue!90!black},
    citecolor={blue!90!black},
    urlcolor={blue}
}

\usepackage[T1]{fontenc}
\usepackage{ae,aecompl}

\usepackage{graphicx}
\usepackage{amsmath}
\usepackage{amssymb}
\usepackage{multirow,bigdelim}
\usepackage{mathtools}
\usepackage{siunitx}
\newcommand{\tred}{\textcolor{red}}
\newcommand{\benj}{\textcolor{black}}

\newcommand{\wgx}{$w_{\rm{g}\times}\,$}
\newcommand{\wgp}{$w_{\rm{g+}}\,$}
\newcommand{\mpc}{\,h^{-1}{\rm{Mpc}}}
\newcommand{\rp}{$r_{p}$}
\newcommand{\ssz}{\scriptsize}
\newcommand{\Aia}{A_{\textrm{\ssz{IA}}}}
\newcommand{\Aiab}{A_{\textrm{\ssz{IA}}}^{\textrm{B}}}
\newcommand{\Aiar}{A_{\textrm{\ssz{IA}}}^{\textrm{R}}}

\usepackage{bm}
\usepackage{soul}
\usepackage{comment}
\numberwithin{equation}{section}

\usepackage{etoolbox}
\makeatletter
\patchcmd\@combinedblfloats{\box\@outputbox}{\unvbox\@outputbox}{}{%
  \errmessage{\noexpand\@combinedblfloats could not be patched}%
}%
\makeatother

\begin{document}

\title{KiDS$+$GAMA: Intrinsic alignment model constraints for current and future weak lensing cosmology}
\titlerunning{KiDS$+$GAMA Intrinsic Alignments}

\author{
Harry Johnston\inst{1}\thanks{\emph{E-mail:} hj@star.ucl.ac.uk},  
Christos Georgiou\inst{2},  
Benjamin Joachimi\inst{1},  
Henk Hoekstra\inst{2},
Nora Elisa Chisari\inst{3},
Daniel Farrow\inst{4},
Maria Cristina Fortuna\inst{2}, 
Catherine Heymans\inst{5},
Shahab Joudaki\inst{3},
Konrad Kuijken\inst{2} and
Angus Wright\inst{6}
}

\authorrunning{H. Johnston et al. }

\institute{Department of Physics and Astronomy, University College London, Gower Street, London WC1E 6BT, UK\and
Leiden Observatory, Leiden University, PO Box 9513, Leiden, NL-2300 RA, the Netherlands\and
Department of Physics, University of Oxford, Keble Road, Oxford, OX1 3RH, UK\and
Max-Planck-Institut f\"{u}r extraterrestrische Physik, Postfach 1312 Giessenbachstrasse, D-85741 Garching, Germany\and
Scottish Universities Physics Alliance, Institute for Astronomy, University of Edinburgh, Blackford Hill, Edinburgh EH9 3HJ, UK\and
Argelander-Institut f\"{u}r Astronomie, Universit\"{a}t Bonn, Auf dem H\"{u}gel 71, 53121 Bonn, Germany}
\date{Accepted XXX. Received YYY; in original form ZZZ}

\label{firstpage}
\makeatletter
\renewcommand*\aa@pageof{, page \thepage{} of \pageref*{LastPage}}
\makeatother

\abstract{
We directly constrain the non-linear alignment (NLA) model of intrinsic galaxy alignments, analysing the most representative and complete flux-limited sample of spectroscopic galaxies available for cosmic shear surveys.  We measure the projected galaxy position-intrinsic shear correlations and the projected galaxy clustering signal using high-resolution imaging from the Kilo Degree Survey (KiDS) overlapping with the GAMA spectroscopic survey, and data from the Sloan Digital Sky Survey. Separating samples by colour, we make no significant detection of blue galaxy alignments, constraining the blue galaxy NLA amplitude $\Aiab=0.21^{+0.37}_{-0.36}$ to be consistent with zero. We make robust detections ($\sim9\sigma$) for red galaxies, with $\Aiar=3.18^{+0.47}_{-0.46}$, corresponding to a net radial alignment with the galaxy density field, and we find no evidence for any scaling of alignments with galaxy luminosity. We provide informative priors for current and future weak lensing surveys, an improvement over de facto wide priors that allow for unrealistic levels of intrinsic alignment contamination. For a colour-split cosmic shear analysis of the final KiDS survey area, we forecast that our priors will improve the constraining power on $S_{8}$ and the dark energy equation of state $w_{0}$, by up to $62\%$ and $51\%$, respectively. Our results indicate, however, that the modelling of red/blue-split galaxy alignments may be insufficient to describe samples with variable central/satellite galaxy fractions. 
}

\maketitle



\section{INTRODUCTION}

Light travelling towards Earth passes through the inhomogeneous universe, and consequent tidal gravitational field. In accordance with General Relativity, the light is differentially deflected, producing coherent distortions in the apparent shapes of source galaxies. This weak cosmological lensing -- or cosmic shear -- signal encodes information pertaining to the total matter distribution, universal geometry, and cosmic expansion and acceleration, as each evolves with redshift. Thus cosmic shear is one of the vital probes in the challenge to de-shroud the dark energy and dark matter species of the $\Lambda$CDM cosmological model.

Since its first detections around the turn of the century (\citealt{Bacon2000}, \citealt{Kaiser2000}, \citealt{Wittman2000}, \citealt{VanWaerbeke2000}), cosmic shear has matured into a powerful tool for cosmology (\citealt{Heymans2013}, \citealt{Jee2016}, \citealt{Hildebrandt2018}, \citealt{Kohlinger2017}, \citealt{Troxel2017}, \citealt{Hikage2018}), been combined with complementary probes to great effect (\citealt{DESCollaboration2017}, \citealt{Joudaki2017},  \citealt{VanUitert2018}) and formed the basis of design for many next-generation wide-field sky surveys (LSST; \citealt{LSSTScienceCollaboration2009}, Euclid; \citealt{Laureijs2011}, WFIRST; \citealt{Spergel2013}).

The primary astrophysical systematic effect for cosmic shear is the \emph{intrinsic alignment} of galaxies (\citealt*{Heavens2000}, \citealt{Croft2000}, \citealt{Catelan2001}, \citealt{Hirata2004a}). Cosmic shear relies upon picking up coherent, percent-level shape distortions (shears) over a statistical ensemble of galaxies. However, galaxies may interact with the gravitational field during formation, becoming aligned with their local environment. The same environment/structure also contributes to the lensing distortions observed in background galaxies. Both processes contaminate cosmic shear signals by sourcing non-random shear correlations in imaging data; between the intrinsic shapes of locally aligned galaxies (II), and between those intrinsic shapes and the gravitational shear field (GI). II correlations are restricted to physically close pairs, and are subdominant to the latter GI term, which can operate over wide separations in redshift, posing a greater threat of contamination for deep cosmic shear studies.

Tidal alignments, as they apply to galaxies, are thought to manifest through two principal mechanisms; galaxy halos are tidally (i) \emph{stretched} (see \citealt{Catelan2001}), and (ii) \emph{torqued} (see \citealt{Schaefer2008} for a review of the latter) by the interaction of the tidal shear quadrupole with the moment of inertia of the halo. Pressure-supported, red elliptical galaxies equilibrate their stellar distributions according to the ellipsoidal halo potential. Rotationally supported, blue spirals align their spin axes with the halo angular momentum (see \citealt{Kiessling2015}). Each type is thus imprinted with the alignments of the halo. The former effect is linear, and the latter quadratic in the matter density contrast, suggesting strong tidal alignment of blue galaxy \emph{spin axes} at small scales, which quickly dissipate with increasing separation. These contrast with the further-reaching \emph{shape} alignments of red galaxies. Both types of alignments should be stronger around more pronounced peaks of the matter distribution \citep{Piras2017}. 

This picture is supported by observations; many studies show strong alignments out to $100\mpc$ for SDSS galaxies, with luminous red galaxies (LRGs) and bright subsamples showing the largest alignment amplitudes (\citealt{Mandelbaum2006}, \citealt{Hirata2007}, \citealt{Joachimi2011}, \citealt{Li2013}, \citealt{Singh2015}). Significant ($\geqslant{}3\sigma{}$) alignments of nearby spiral galaxy spin axes, with reconstructed tidal fields, have been reported on scales $\lesssim{3}\, h^{-1}\rm{Mpc}$ (\citealt*{Lee2002}, \citealt*{Lee2007},  \citealt{Lee2011}), but attempted measurements of large-scale intrinsic ellipticity correlations of spirals have thus far been consistent with zero (\citealt{Hirata2007}, \citealt{Mandelbaum2011}, \citealt{Tonegawa2017}). Hydrodynamical simulations corroborate these observational findings for red galaxies, but exhibit disagreements as to the form and amplitude of blue galaxy alignments (\citealt{Chisari2015}, \citealt{Velliscig2015}, \citealt{Tenneti2016}, \citealt{Hilbert2017}).


The risk of shear contamination by intrinsic alignment (IA) of galaxies, and the associated threat posed to cosmological parameter inference, has long been known (\citealt*{Heavens2000}, \citealt{Heymans2004}, \citealt{Hirata2004a}). Much work has been devoted to measuring the strength of IA and forecasting its impact under various scenarios of modelling or lack thereof (\citealt{Joachimi2010}, \citealt{Joachimi2011}, \citealt{Kirk2012}, \citealt*{Krause2016}, \citealt{Blazek2017}). Broadly summarised, the findings suggest (i) significant biasing of cosmological parameters if IA are not accounted for; (ii) IA mitigation schemes, involving nuisance parameters for marginalisation, which will degrade cosmological constraints but can effectively mitigate biasing of parameter inference; (iii) joint analyses of shear probes with positional information and cross-correlations, to aid with degeneracy-breaking and self-calibration of IA models; (iv) the need for increasingly detailed modelling of IA -- particularly with respect to non-linearities -- accompanied by simulations (for model-testing and predictions) and observational constraints upon IA parameters over a long redshift baseline.

Recent, dedicated studies of cosmic shear have allowed for the effects of intrinsic alignments with nuisance parameterisations (\citealt{Heymans2013}, \citealt{Jee2016}, \citealt{Joudaki2016}, \citealt{Hildebrandt2018}, \citealt{Troxel2017}, \citealt{Samuroff2018}). The currently preferred models, with wide prior ranges, wield great power to modulate lensing observables -- this has resulted in heavy degradation of cosmological constraining power. Moreover, we cannot be certain that other systematic effects, known (e.g. photo-$z$ errors -- see \citealt{Efstathiou2018}, \citealt{VanUitert2018}) or otherwise, are not leaking into the IA parameterisations. Informative priors for the models are the first step to assuaging these concerns, and they must be derived from galaxy samples \emph{representative} of cosmic shear datasets.

This work aims to motivate such a prior for current and future studies by constraining the alignment amplitudes exhibited by the flux-limited GAMA spectroscopic sample \citep{Driver2011a}, with high-resolution KiDS \citep{DeJong2013} imaging and shapes. We supplement our GAMA data with galaxies from the SDSS Main sample (\citealt{York2000}, \citealt{Strauss2002}) -- the only other readily available, wide-area, flux-limited, spectroscopic dataset. This study is made unique by the lack of selection \emph{a priori}, high completeness ($>98\%$) and spectroscopic redshifts of GAMA and SDSS Main, and so yields a set of constraints which are uniquely instructive for future shear studies. With the aforementioned dependencies of alignments in mind, we also split our samples by colour and redshift, and fit to them with colour-specific parameters, in an effort to more comprehensively describe the contributions of the two galaxy populations.

We measure galaxy position-intrinsic shear correlations in our samples, using galaxies as a proxy to the total matter density field and measuring their tendency to align with that field over a range of scales. We simultaneously fit to clustering measurements in the same samples for self-calibration of the galaxy bias, elsewise degenerate with the intrinsic alignment amplitude. We fit to our signals with the non-linear alignment (NLA) model (\citealt{Hirata2004a}, \citealt{Bridle2007}), with and without a luminosity power-law. We forecast, via Fisher matrix analysis, the improvement in cosmological parameter constraints for a finished KiDS survey, when adopting our derived IA constraints as informative priors.


The structure of this paper is as follows; we describe our galaxy survey data in Section \ref{sec:data}, along with our measurement pipeline. Section \ref{sec:modelling} details our models and methods of fitting, and we summarise the results of fitting in Section \ref{sec:results}. Section \ref{sec:forecasting} outlines our forecasting for a future shear analysis, and our concluding remarks are presented in Section \ref{sec:conclusions}.

Throughout our intrinsic alignment analysis, we work with rest-frame AB magnitudes, $k$-corrected to $z=0$, and assume a flat $\Lambda$CDM cosmology with $\Omega_{m}=0.25$, $h=0.7$, $\Omega_{b}=0.044$, $n_{s}=0.95$, $\sigma_{8}=0.8$, $w_{0}=-1$ and $w_{a}=0$. This is the cosmology adopted by the MICE\footnote{Publicly available through CosmoHub: \url{http://cosmohub.pic.es}} simulations, whose mocks we make use of in our analysis (see Appendix \ref{sec:app_mice_errors}). It is also similar to that assumed by \cite{Joachimi2011}, allowing for direct comparison of intrinsic alignment constraints.

\section{DATA}
\label{sec:data}

\begin{figure*}
	\includegraphics[width=\textwidth]{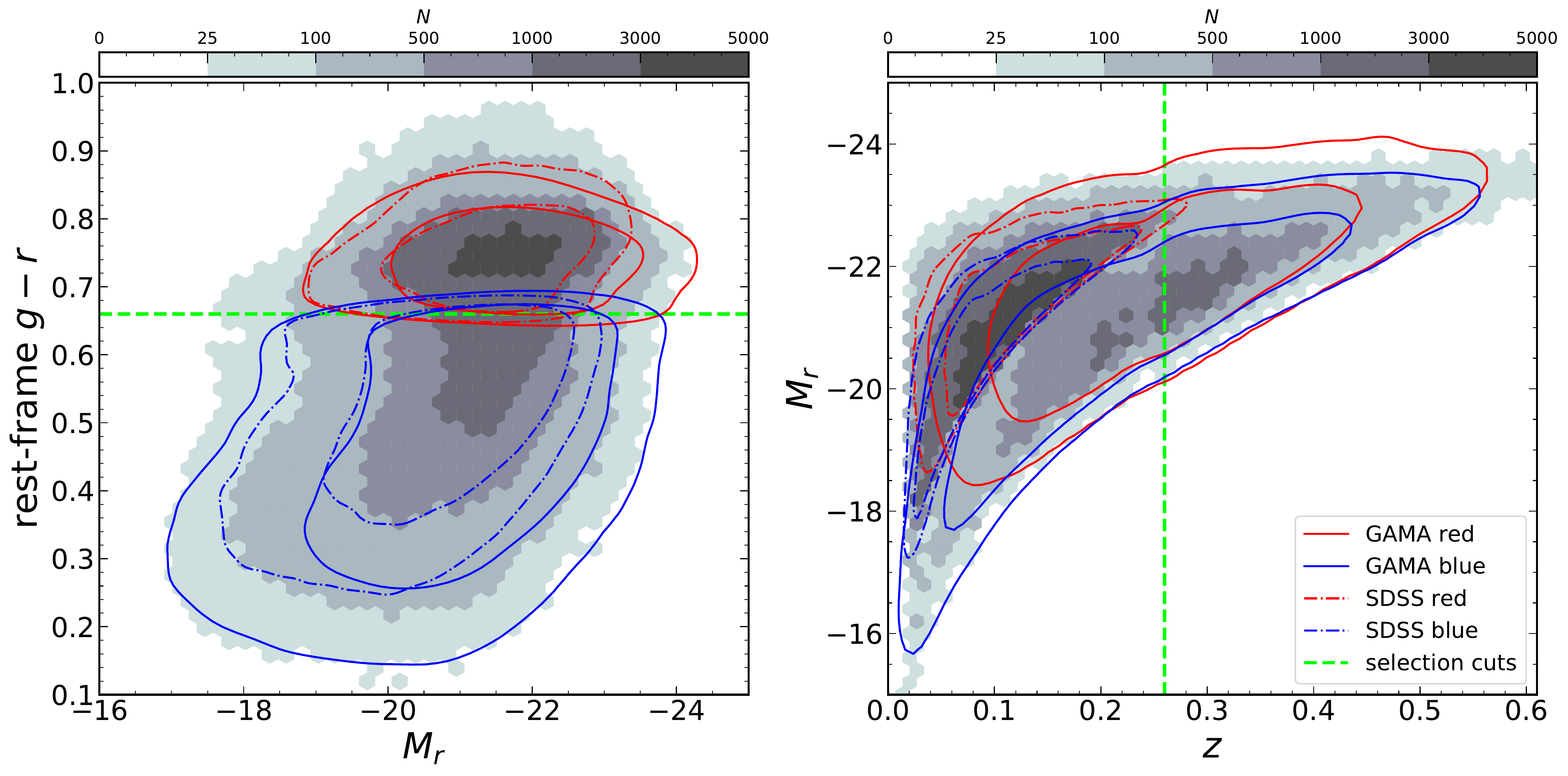}
    \caption{\textit{Left:} Galaxy rest-frame colour-magnitude diagram, where we choose a cut in $g-r$ to isolate the red sequence in GAMA and SDSS. \textit{Right:} Sample absolute $r$-band magnitude-redshift diagram. The total distribution of GAMA and SDSS galaxies is shown, binned in hexagonal cells with a colour scale corresponding to the counts in cells. Coloured contours indicate $75\%$ and $95\%$ of galaxies in a sample. Colour/redshift cuts are shown as dashed green lines, and the apparent leakage of contours is due to the grid-size used in kernel density estimation.}
    \label{fig:sample_spaces}
\end{figure*}

\subsection{KiDS$\,+$GAMA}
\label{sec:data_kids_gama}

The ongoing Kilo Degree Survey (KiDS; \citealt{DeJong2017}) is a wide-field optical imaging survey, taking data in four passbands (\textit{ugri}) with the OmegaCAM camera at the VLT Survey Telescope (VST). The VST-OmegaCAM system is optimised for producing 1$\,\,\rm{deg}^{2}$ images of exceptional quality, facilitating accurate galaxy shape measurements for the primary science driver of KiDS: weak lensing studies.

KiDS aims to image $1350\,\,\rm{deg}^{2}$ of sky in 2 rougly equal-sized strips. KiDS-North, centered on the celestial equator, shares complete overlap with the Galaxy and Mass Assembly (GAMA; \citealt{Driver2011a}) equatorial fields -- a total $180 \,\rm{deg}^{2}$, split equally between G9, G12 and G15. GAMA is a now-complete spectroscopic survey which operated on the Anglo-Australian Telescope, with the AAOmega spectrograph. GAMA galaxies possess thoroughly tested spectroscopic redshifts and are highly complete ($>98\%$) in the $r$-band limit $r<19.8$.

Our KiDS$+$GAMA dataset consists of the final release \citep{Liske2015}, equatorial GAMA spectroscopic sample, with shapes measured from KiDS-450\footnote{\url{http://kids.strw.leidenuniv.nl/DR3/}} imaging. Galaxy shapes are measured from $r$-band images, for which the best dark-time seeing conditions are reserved in KiDS. \cite{Singh2016b} analysed the SDSS-III BOSS LOWZ luminous red galaxy (LRG) sample \citep{Alam2015} with different shape measurement methods, finding variability in ellipticities, intrinsic alignment conclusions and the impacts of observational systematics. The connection between such variabilities and the radial weighting employed in shape estimation is explored in our companion paper: \cite{Georgiou2018}.

We measure shapes using the moments-based DEIMOS (DEconvolution In MOments Space) method introduced by \cite{Melchior2011}. We briefly describe the DEIMOS method here,
as applied to KiDS$+$GAMA, and refer the reader to \cite{Georgiou2018} for details of the production of our ellipticity catalogue. The moments \tens{Q} of the distribution of brightness (flux) $G(\vec{x})$ across an image, where $\vec{x}=(x\,,y)$ is a coordinate vector, are expressed in Cartesian coordinates as

\begin{equation}
	\tens{Q}_{ij} = \int G(\vec{x})\, x^{i} y^{j} \, {\rm{d}}x\, {\rm{d}}y \quad ,
	\label{eq:moments}
\end{equation}
where $n=i+j$ gives the order of the moment. One recovers the complex ellipticity of an object from $2^{\textrm{nd}}$-order brightness moments as

\begin{equation}
	\label{eq:ellipticity}
	\epsilon \equiv \epsilon_{1} + \rm{i}\epsilon_{2} = \frac{ \tens{Q}_{20} - \tens{Q}_{02} + 2\rm{i}\,\tens{Q}_{11} }{ \tens{Q}_{20} + \tens{Q}_{02} + 2\, \sqrt{\tens{Q}_{20}\, \tens{Q}_{02} - \tens{Q}_{11}^{2}} } \quad ,
\end{equation}
which relates to the semi-major $a$ and semi-minor $b$ axes as $|\epsilon| = (a-b)(a+b)^{-1}$.

Observed galaxy flux profiles $G^{\vec{*}}(\vec{x})$ are distorted by convolution with the point spread function $P(\vec{x})$ -- determined by observing conditions, telescope optics and detector properties, the PSF describes the blurring of point-like sources in imaging. The PSF-convolved flux profile is

\begin{equation}
	G^{\vec{*}}(\vec{x}) = \int G(\vec{x}')\, P(\vec{x} - \vec{x}')\, {\rm{d}}\vec{x}' \quad .
\end{equation}
\cite{Melchior2011} transform the flux profile into Fourier space and show, with the convolution theorem, that the moments of the observed flux distribution $\tens{Q}^{\vec{*}}_{ij}$ are

\begin{equation}
	\tens{Q}^{\vec{*}}_{ij} = \sum^{i}_{k} \sum^{j}_{l} \, \begin{pmatrix} \, i\, \\ k \end{pmatrix} \begin{pmatrix} j \\ \, l\, \end{pmatrix} \, \tens{Q}_{kl} \{P\}_{i-k,j-l} \quad ,
\end{equation}
where $\{P\}_{ij}$ denotes the moments of the PSF. Thus the $n^{\textrm{th}}$-order deconvolved moments $\tens{Q}_{ij}$ of the image can be recovered from the image- and PSF-moments up to the same order. In practice, one must also account for noise in the image, from sky background, pixel noise etc. The pixel signal-to-noise ratio (SNR) is lowest at large distances from the galaxy centroid, which would tend to dominate the measurement of $2^{\textrm{nd}}$-order brightness moments (Eq. \ref{eq:moments}). We suppress pixel noise using Gaussian elliptical weight functions $W(\vec{x})$, and recover an approximation to the unweighted brightness profile by computing a truncated Taylor expansion of $W^{-1}(\vec{x})$ \citep[see][]{Georgiou2018}.

Galaxy shapes can be obscured by overlapping objects in images. These shapes can still be measured by applying masks to the nuisance objects, but the loss of information could have an impact upon the quality of the shape measurement. We verify that excluding blended galaxies -- where isophotal radii overlap -- does not significantly change our measurements of alignment correlation functions, and continue to include these galaxies in our analysis. We refer the reader to \cite{Georgiou2018} for further details on our use of weight functions and associated bias considerations, deblending, and any other details of the shape measurements.

We choose a rest-frame colour cut of $g-r>0.66$ on inspection of the colour-$r$-band absolute magnitude diagram, in order to cleanly isolate the red sequence (Figure \ref{fig:sample_spaces}), and we define 2 redshift bins with edges $[\,0.02\,,0.26\,,0.5\,]$ (see Appendix \ref{sec:app_covariances} for more detail on this choice). These cuts yield colour/redshift samples (Z1B, Z1R, Z2B, Z2R) of roughly equal size, and we apply the same colour cut to SDSS galaxies. For measuring position-intrinsic shear correlations in each sample, we define a `shapes' subset of galaxies residing in unmasked\footnote{Our mask excludes galaxies in pixels subject to readout spikes, saturation cores, diffraction spikes, primary halos of foreground objects, bad pixels and manually masked regions.} pixels (for details of the masking procedure, see \cite{Kuijken2015} We further exclude any galaxies flagged as having a bad shape measurement (see \citealt{Georgiou2018}), and correlate the remaining ($\sim85-87\%$) shapes against the positions of all galaxies in the same colour/redshift bin -- the `density' sample. We also measure correlations against randomly distributed points, using random catalogues specifically designed for GAMA \citep{Farrow2015}, and randomly downsampled to retain at least $10\times$ the number of galaxies in a corresponding galaxy sample. Where used in additional, demonstrative sample selections, stellar-mass estimates for GAMA galaxies are taken from {\tt StellarMassesLambdarv20} \citep{Wright2017}.

\subsection{SDSS Main}

The Sloan Digital Sky Survey (SDSS; \citealt{York2000}) imaged about $\pi$ steradians of the sky, drift-scanning in five bands ($ugriz$), with the purpose-built, wide-field SDSS photometric camera \citep{Gunn1998}. Of the $\sim{1}$ million objects followed up spectroscopically, the Main galaxy sample \citep{Strauss2002} was designed to be flux-limited and highly complete ($>99\%$) to $r<17.77$, thus forming a complementary dataset to GAMA, shallower and over a wider area of $\sim{3340}\,\rm{deg}^{2}$. These are the same SDSS Main samples measured for IA by \cite{Mandelbaum2006}, \cite{Hirata2007} and \cite{Joachimi2011}, where the latter two works also included LRG-selected samples in their analysis. We make no magnitude selections, and employ a different colour-cut in our analysis, hence we re-measure the alignment signals. We use PSF-corrected ellipticity measurements made by \cite{Mandelbaum2005} with the {\sc{Reglens}} pipeline -- {\sc{Reglens}} measures galaxy shapes via `re-Gaussianisation' \citep{Hirata2003a}. This is a moments-based method, which assumes Gaussianity in the PSF and galaxy profiles, treating non-Gaussianities with perturbative corrections. We refer the reader to \cite{Hirata2003a}; \cite{Mandelbaum2005} for further details.

We define red and blue SDSS samples (SR, SB) with the same rest-frame cut at $g-r=0.66$. The SDSS density samples retain galaxies with bad shapes flags, which are excluded from the shapes samples. Figure \ref{fig:sample_spaces} illustrates the colour-redshift-magnitude spaces of our selected samples, which are detailed in Table \ref{tab:sample_details}.

\begin{table}
	\small
	\centering
	\caption{Details of our density (bracketed numbers) and intrinsic shape field tracer samples, composed of GAMA and SDSS galaxies split by redshift and/or colour. $L_{\rm{piv}}\sim\num{4.6e10}L_{\sun}$ corresponds to an absolute $r$-band magnitude of $-22$. For purposes of clustering covariance estimation (see Appendix \ref{sec:app_mice_errors}), we impose a faint limit $M_{r}\leqslant-18.9$ on our GAMA density samples -- hence Z1B has fewer density galaxies than shapes.}
	\label{tab:sample_details}
   	\def\arraystretch{1.2}
	\begin{tabular}{lccc} 
		\hline
		 Sample & $\langle{}z\rangle$ & $\langle{}L/L_{\rm{piv}}\rangle$ & $N$ shapes (density)\\
		\hline
		\hline
		GAMA $z>0.26$, & \multirow{2}{*}{0.33} & \multirow{2}{*}{1.06} & \multirow{2}{*}{31447 (36791)}\\
        blue (Z2B) & & & \\
		\hline
		GAMA $z<0.26$, & \multirow{2}{*}{0.15} & \multirow{2}{*}{0.21} & \multirow{2}{*}{60634 (52273)}\\
        blue (Z1B) & & & \\
		\hline
		SDSS blue (SB) & 0.09 & 0.14 & 110557 (114054)\\
		\hline
		GAMA $z>0.26$, & \multirow{2}{*}{0.33} & \multirow{2}{*}{1.47} & \multirow{2}{*}{31368 (36087)}\\
        red (Z2R) & & & \\
		\hline
		GAMA $z<0.26$, & \multirow{2}{*}{0.17} & \multirow{2}{*}{0.50} & \multirow{2}{*}{38011 (42078)}\\
        red (Z1R) & & & \\
		\hline
		SDSS red (SR) & 0.12 & 0.29 & 166198 (171565)\\
		\hline
	\end{tabular}
\end{table}

\subsection{Estimators}
\label{sec:estimators}

We adapt the notation of \cite{Schneider2002a}, defining a bin filter $\Delta_{r_{p},\Pi}(\vec{x}) = 1$ for a pair separation vector $\vec{x}=(x_{\parallel}\,,\,x_{\perp})$ where the (absolute) comoving radial component $x_{\parallel}$ is less than the maximum under consideration $\Pi_{\rm{max}}$, and the comoving transverse component $x_{\perp}$ satisfies $r_{p} / 10^{\Delta\textrm{log}(r_{p})/2} < x_{\perp} \leqslant  r_{p} \times 10^{\Delta\textrm{log}(r_{p})/2}$ for a transverse bin centred on $r_{p}$ (log-space bin width $\Delta\textrm{log}(r_{p})$ is constant). For any other separation vector, $\Delta_{r_{p},\Pi}(\vec{x}) = 0$. We adopt the estimator defined by \cite{Mandelbaum2006}\footnote{For ease of computation, we actually normalise by the density-randoms vs. shapes paircount $N_{rs}(r_{p}, \Pi)$, as opposed to the  density-randoms vs. shapes-randoms paircount $N_{rr_{s}}(r_{p}, \Pi)$. We verify that resulting estimates differ negligibly with respect to the noise level.}, and given as

\begin{equation}
 \label{eq:m06_estimator}
 \begin{split}
	& \hat{\xi}_{\rm{g}+}(r_{p}, \Pi) \, = \, \frac{1}{\, N_{rr_{s}} \, (r_{p},\Pi)} \, \, \times \\
    & \quad \quad \left\{\sum_{sd} \,\gamma_{+,sd} \, \Delta_{r_{p},\Pi}(\bm{x}_{s}-\bm{x}_{d})\right. 
	\, \left. - \, \, \sum_{sr} \,\gamma_{+,sr} \, \Delta_{r_{p},\Pi}(\bm{x}_{s}-\bm{x}_{r})\right\} \quad , 
 \end{split}
\end{equation}
where we use subscripts to denote and index shape ($s$), density ($d$) and density-/shapes-random ($r\,,r_{s}$) galaxy samples, and

\begin{equation}
	N_{ij}(r_{p},\Pi) \, = \, \sum_{ij} \, \Delta_{r_{p},\Pi}(\vec{x}_{i} - \vec{x}_{j}) \quad ,
\end{equation}
gives the paircount between samples $i$ and $j$, which is then normalised according to the relative sample populations\footnote{The normalisation is by $n_{i}n_{j}$ when $i\neq{j}$, or else by $n_{i}(n_{i} - 1)$, i.e. the total possible number of galaxy pairs with unlimited separations.}. The tangential shear component\footnote{In practice, one could affix weights $w_{s}$ to the shear components, to allow for down-weighting of noisier shapes -- we do not apply any weights in our analysis (nor do previous direct-measurement studies of IA), as our use of elliptical weight functions in shape estimation poses problems for the estimation of ellipticity errors (see Section 2.3 of \citealt{Georgiou2018}).} $\gamma_{+,ij}$ of a galaxy $i$ relative to the vector connecting it to a galaxy $j$ is given as

\begin{equation}
	\gamma_{+,ij} \, = \, \frac{1}{\mathcal{R}} \, \Re{\rm{e}} \left[\epsilon_{i} \, {\rm{exp}}\left(-2{\rm{i}} \varphi_{ij} \right) \right] \quad ,
	\label{eq:epsilon_+}
\end{equation}
where, for galaxy $i$, the ellipticity $\epsilon_{i} = \epsilon_{i1} + {\rm{i}}\epsilon_{i2}$ (see Section \ref{sec:data_kids_gama}) and $\varphi_{ij}$ is the polar angle of the pair separation vector. Note that the sign convention here is $\gamma_{+}>0$ for \emph{radial} alignments, in contrast with the standard for galaxy-galaxy lensing. The shear responsivity $\mathcal{R}\approx 1 - \sigma_{\epsilon}^{2}$ in Eq. \ref{eq:epsilon_+} quantifies the response of galaxy ellipticities to gravitational shearing, for a given galaxy sample. The responsivity is doubled when ellipticities are measured via polarisation \citep[see][]{Mandelbaum2006}, as is the case for our SDSS samples. The resulting shear corrections are then $\lesssim8\%$ for GAMA, and $\lesssim35\%$ for SDSS, respectively.

We consider our measurements in line-of-sight projection

\begin{equation}
	w_{\rm{g}+}(r_{p}) = \int^{\Pi_{\rm{max}}}_{-\Pi_{\rm{max}}} \xi_{\rm{g}+}(r_{p},\Pi) \, {\rm{d}}\Pi \quad ,
	\label{eq:wgp_definition}
\end{equation}
thus compressing the measurement into fewer data points, with generally higher signal-to-noise ratios (S/N).

We test for alignment systematics by measuring (i) the position-intrinsic correlation cross-component \wgx (replacing $\gamma_{+}$ with $\gamma_{\times}$ in Eq. \ref{eq:m06_estimator}, where $\gamma_{\times}$ is the imaginary analogue to Eq. \ref{eq:epsilon_+}; equivalent to $\gamma_{+}$ after a 45 degree rotation of the ellipticity), which must vanish on average since galaxy formation does not break parity. We also measure (ii) \wgp for galaxy pairs with large line-of-sight separations $60\leqslant|\Pi|\leqslant90\mpc$. Spectroscopic redshifts allow us to choose a narrower range for this test, relative to previous works (e.g. \citealt{Joachimi2011}), starting at $60\mpc$ given recent detections of alignments on large transverse scales \citep{Singh2015}. One expects astrophysically induced alignment signals to be dominated by short-range correlations, and consistent with zero over much larger scales, providing the second, ``large-$\Pi$" systematics test.

We measure galaxy clustering with the standard \cite{Landy1993} estimator

\begin{equation}
	\xi_{\rm{gg}}(r_{p},\Pi) = \frac{N_{dd} - 2N_{dr} + N_{rr}}{N_{rr}} \quad ,
	\label{eq:xi_gg}
\end{equation}
where the $r_{p}\,, \Pi$ binning of paircounts is implicit. Eq. \ref{eq:xi_gg} is well known to improve the bias and covariance properties of the galaxy auto-correlation through subtraction of the random field from the density field, and this concept carries over to our alignment estimator (Eq. \ref{eq:m06_estimator}).

\cite{Singh2016} demonstrated that the subtraction of the galaxy-galaxy lensing signal measured around random points (i.e. a randomly distributed lens sample) also holds advantages in reducing the impact of systematics and correlated shape noise, especially on large transverse scales. This is done in the context of galaxy-galaxy lensing; long-range lens clustering introduces noise through lensed, and therefore correlated, background shapes. The GI analogy would suppose that \emph{intrinsic} shears of the shape sample are correlated over super-sample scales -- e.g. galaxies aligning with filaments/knots etc. This correlated shape noise would show in the random-intrinsic correlation, and be subtracted by our estimator (Eq. \ref{eq:m06_estimator}).

We compute the total projected correlation functions by summing over line-of-sight separations $-60\leqslant\Pi\leqslant+60\mpc$, in bins of $\Delta\Pi = 4\mpc$ (Eq. \ref{eq:wgp_definition} \& analogous for $w_{\rm{gg}}$) and consider the results in 11 log-spaced bins between $0.1\leqslant r_{p}\leqslant60\mpc$.

We compute all intrinsic alignment correlations using our own code, and make use of the public {\sc{swot}}\footnote{\url{https://github.com/jcoupon/swot}} \citep{Coupon2012} kd-tree code for clustering correlations, which we verify against our own (brute-force) code and against external clustering measurements in GAMA \citep{Farrow2015}.

\subsection{Covariances}

We estimate signal covariances with delete-one jackknife methods, which we describe briefly here, referring the reader to Appendix \ref{sec:app_covariances} for more detail.

Jackknife samples are defined by the consecutive exclusions and replacements of many `patches' within the survey footprint, such that each sample constitutes most of the galaxy data. The covariance is thus estimated by considering the deviation from the mean signal upon removal of independent subsets of the data. Each subset must then correctly and independently sample the signal of interest; each patch must be larger than the largest scales under examination. Simultaneously, the number of patches must be much greater than the size of the data vector, or else estimates of the inverse covariance will suffer from excessive noise. Attempting to satisfy both requirements, we implement a $3\rm{D}$ jackknife routine, slicing patches in redshift and multiplying the available number of independent subsets. We remain, however, unable to reliably sample large pair separations at low-redshift in GAMA (see Figure \ref{fig:patchsizes}), thus we discard the largest scales ($\sim40 - 60\mpc$) for GAMA samples with a significant proportion of low-redshift galaxies -- see Appendix \ref{sec:app_covariances} for more detail, and for assessments of the jackknife performance.

\section{MODELLING}
\label{sec:modelling}

We observe the weak lensing angular power spectrum as the sum of shear-shear (GG), intrinsic-intrinsic (II) and shear-intrinsic (GI) contributions, such that

\begin{equation}
	C_{ij}(\ell) = C^{\rm{GG}}_{ij}(\ell) + C^{\rm{II}}_{ij}(\ell) + C^{\rm{GI}}_{ij}(\ell) \quad ,
	\label{eq:shearspectrum}
\end{equation}
for correlations between samples $i$ and $j$. The cosmic shear GG term encodes the average coherent gravitational shearing of galaxies' light by structure along the line-of-sight, and is the statistic of interest for cosmological analyses. Intrinsically correlated orientations of galaxies result in the extra intrinsic shear correlation II and interference GI terms. These angular power spectra are theoretically determined for a flat universe, by Limber projection of the matter $P_{\delta}$, intrinsic $P_{\rm{II}}$ and matter-intrinsic $P_{\delta\rm{I}}$ power spectra, as

\begin{equation}
	C^{\rm{GG}}_{ij} (\ell) = \int^{\chi_{\rm{h}}}_{0} {\rm{d}}\chi \, \frac{q^{\, (i)}(\chi)q^{\, (j)}(\chi)}{\chi^{2}} \, P_{\delta}\left(\frac{\ell}{\chi}, \chi\right) 
	\label{eq:C_GG}
\end{equation}

\begin{equation}
	C^{\rm{II}}_{ij} (\ell) = \int^{\chi_{\rm{h}}}_{0} {\rm{d}}\chi \, \frac{p^{\, (i)}(\chi)p^{\, (j)}(\chi)}{\chi^{2}} \, P_{\rm{II}}\left(\frac{\ell}{\chi}, \chi\right) 
	\label{eq:C_II}
\end{equation}

\begin{equation}
	C^{\rm{GI}}_{ij} (\ell) = \int^{\chi_{\rm{h}}}_{0} {\rm{d}}\chi \, \frac{q^{\, (i)}(\chi)p^{\, (j)}(\chi) + p^{\, (i)}(\chi)q^{\, (j)}(\chi)}{\chi^{2}} \, P_{\delta\rm{I}}\left(\frac{\ell}{\chi}, \chi\right) \quad ,
	\label{eq:C_GI}
\end{equation}
each weighted by an efficiency kernel describing the coincidence of sample redshift (comoving distance) distributions $p(\chi)$ and/or lensing efficiencies $q(\chi)$, where $p(\chi)\, {\rm{d}}\chi=p(z)\, {\rm{d}}z$ and

\begin{equation}
	q(\chi) = \frac{3H_{0}^{2}\Omega_{\rm{m}}}{2c^{2}} \int^{\chi_{\rm{h}}}_{\chi} {\rm{d}}\chi{'} \, p(\chi{'}) \, \frac{\chi{'}-\chi}{\chi{'}} \quad ,
	\label{lens_eff}
\end{equation}
for present-day Hubble parameter $H_{0}$, matter energy-density fraction $\Omega_{\rm{m}}$ and comoving distances $\chi$, with $\chi_{\rm{h}}$ denoting the horizon distance. The redshift distributions of our galaxy samples are shown in Figure \ref{fig:patchsizes} as dashed red/blue histograms.

We constrain models for $P_{\delta\rm{I}}$ by fitting to the real-space alignment and clustering correlation functions described in Section \ref{sec:estimators}.

\subsection{Tidal alignments}
\label{sec:nla}

The linear alignment (LA) model assumes a linear relation between the tidal shearing of galaxies and the gravitational potential quadrupole at their epoch of formation. This form is motivated as follows: fluctuations in the large-scale potential govern the perturbation of halo ellipticites, within which galaxy ellipticities follow suit. With the large-scale fluctuations necessarily small, higher-order terms dwindle and the intrinsic shearing of galaxies by large-scale structure is thus assumed to be a localised, linear function of the potential. In the simplest case, this leads to intrinsic shear $P_{\rm{II}}$ and cross matter-intrinsic shear $P_{\delta\rm{I}}$ power spectra \citep{Hirata2004a}

\begin{equation}
	P_{\rm{II}}(k, z) = \left(\Aia C_{1} \frac{a^{2} \bar{\rho}(z)}{D(z)}\right)^{2} P_{\delta}(k, z)
	\label{eq:la_model_ii}
\end{equation}
and
\begin{equation}
	P_{\delta\rm{I}}(k, z) = -\Aia C_{1} \frac{a^{2} \bar{\rho}(z)}{D(z)} P_{\delta}(k, z) \quad ,
	\label{eq:la_model}
\end{equation}
respectively, where $\Aia$ is a free, dimensionless amplitude parameter, normalised to unity by the constant ${C_{1}=\num{5e-14}\,M_{\sun}^{-1}h^{-2}\rm{Mpc}^{3}}$ -- this factor is derived by comparing to the work of \cite{Brown2002} who measured II correlations in the low-redshift ($z\sim{0.1}$) SuperCOSMOS survey \citep{Hambly2001}, where cosmic shear is negligible. $\bar{\rho}(z)$ is the mean density of the universe and $D(z)$ is the growth factor.

In the original LA model, $P_{\delta}(k,z)$ is the linear matter power spectrum. \cite{Hirata2007} and \cite{Bridle2007} suggested and implemented a substitution of the non-linear corrected spectrum $P^{\,\rm{nl.}}_{\delta}(k,z)$, birthing the non-linear alignment (NLA) model. Whilst without theoretical motivation, this model was seen to provide a better description of the alignments measured in LRG samples on scales approaching the non-linear. We conduct and present our analysis with both the LA and NLA models, choosing to focus on the NLA given its widespread use in the literature.
Results between the N/LA models will differ only mildly for this work, since meaningful fits of these models must be restricted to quasi-linear scales -- neither model provides a true consideration of non-linear evolution/dependence or of intra-halo baryonic physics (e.g. stellar/AGN feedback). However, the choice of model is expected to levy significant changes in cosmic shear analyses that extract a large fraction of their constraining power from highly non-linear scales. The development of appropriate models for IA remains an active topic of research.


We make fits of the NLA and also a luminosity-dependent analogue, henceforth NLA-$\beta$, including a power-law scaling $\beta$ on the average luminosity $L$ of samples, such that

\begin{equation}
	\Aia \longrightarrow \, \, A_{\beta} \, \left\langle \frac{L}{L_{\rm{piv}}} \right\rangle^{\beta} \quad ,
	\label{eq:nla-beta}
\end{equation}
where $L_{\rm{piv}}\sim\num{4.6e10}L_{\sun}$ is an arbitrary pivot luminosity, corresponding to an absolute $r$-band magnitude of $-22$ (see Table \ref{tab:sample_details}).

We note that fitting linear models to spiral galaxy alignments is at best an approximation to lowest order\footnote{\cite{Hui2008} and \cite{Blazek2017} theorise linear alignment scaling for all galaxies on sufficiently large scales, arising from non-Gaussian structure fluctuations.}, and that next-stage lensing studies should consider splitting the modelling of alignments to include a quadratic alignment prescription for blue galaxies -- such an analysis was recently completed by \cite{Samuroff2018}; applying the mixed alignment model of \cite{Blazek2017} to DESY1 data \citep{DESCollaboration2017}, they find the first marginal evidence for quadratic alignments of both late- and early-type galaxies.

\subsection{Line-of-sight projection}

We project matter and matter-intrinsic power spectra along the line-of-sight using Hankel transformations

\begin{equation}
	w_{\rm{g+}}(r_{p}) = -b_{\rm{g}}\int {\rm{d}}z \, \mathcal{W}(z) \int^{\infty}_{0} \frac{{\rm{d}}k_{\perp}k_{\perp}}{2\pi} J_{2}(k_{\perp}r_{p}) P_{\delta\rm{I}}(k_{\perp}, z)
	\label{eq:g+_hankel}
\end{equation}

\begin{equation}
	w_{\rm{gg}}(r_{p}) = b_{\rm{g}}^{2}\int {\rm{d}}z \, \mathcal{W}(z) \int^{\infty}_{0} \frac{{\rm{d}}k_{\perp}k_{\perp}}{2\pi} J_{0}(k_{\perp}r_{p}) P_{\delta}(k_{\perp}, z) \quad ,
	\label{eq:gg_hankel}
\end{equation}
and jointly model position-intrinsic alignments and clustering, thereby self-calibrating for galaxy bias. $J_{n}$ denotes an $n^{\textrm{th}}$-order Bessel function of the first kind, and $b_{\rm{g}}$ is the linear, assumed scale-independent galaxy bias. The weight function $\mathcal{W}(z)$, as derived by \cite{Mandelbaum2011}, is given by

\begin{equation}
	\mathcal{W}(z) = \frac{p_{i}(z)p_{j}(z)}{\chi^{2}(z)\chi'(z)} \left[ \int {\rm{d}}z \frac{p_{i}(z)p_{j}(z)}{\chi^{2}(z)\chi'(z)} \right]^{-1} \quad ,
\end{equation}
where $p(z)$'s are the normalised redshift probability distributions of the galaxy samples being correlated, i.e. a density and a shapes sample for alignments, or two density samples for clustering. The galaxy samples we analyse in this work are flux-limited, therefore $p(z)$ does not increase as ${\rm{d}}V_{{\rm{com}}}/{\rm{d}}z$ -- the gain in comoving volume with respect to redshift. $\chi(z)\,,\chi'(z)$ are the comoving radial coordinate and its derivative with respect to $z$, such that $\chi^{2}(z)\chi'(z)$ is proportional to ${\rm{d}}V_{{\rm{com}}}/{\rm{d}}z$. Thus, $\mathcal{W}(z)$ is inversely proportional to ${\rm{d}}V_{{\rm{com}}}/{\rm{d}}z$ and acts to down-weight higher redshifts, where flux-limited samples miss faint galaxies.

\subsection{Likelihoods}
	\label{sec:likelihoods}

We constrain the N/LA models, fitting to \wgp and $w_{\rm{gg}}$ (see Section \ref{sec:estimators}) by sampling multi-dimensional parameter posterior distributions, using the {\sc{CosmoSIS}}\footnote{\url{https://bitbucket.org/joezuntz/cosmosis/wiki/Home}} \citep{Zuntz2015} implementation of the affine-invariant {\sc{emcee}} \citep{Foreman-Mackey2012} Monte Carlo Markov Chain sampler. The {\sc{CosmoSIS}} framework supports the  flexible construction of a pipeline to compute theoretical power spectra and other statistics, and to calculate likelihoods against a data vector whilst sampling over parameters. We exclude the first $30\%$ of samples for a burn-in phase.

The non-linear processes unaccounted for by the N/LA models include non-linear density evolution and galaxy biasing, quadratic tidal torquing, and any other higher-order effects contributing to alignment signatures. The galaxy density-weighted sampling of the intrinsic alignment field is included at lowest order in the original derivation by \cite{Hirata2004a}, however \cite*{Blazek2015} highlight additional, linear-scale, galaxy bias-dependent contributions in a perturbative expansion. In light of the models' limitations, and inline with previous analyses, we limit our NLA (and LA) fits to transverse scales above $6\mpc$.

Our parameter vectors for the NLA/NLA-$\beta$ (and LA/-$\beta$) models are then

\begin{equation}
 \begin{array}{l}
	\vec{\lambda}^{\rm{NLA}} = \{\,b_{\rm{g}}\,, \textrm{\small{IC}}\,\}_{i} +  \{\,\Aia\,\}_{\textrm{\ssz{R,B}}} \quad \\[5pt]
	\vec{\lambda}^{{\rm{NLA-}}\beta} = \{\,b_{\rm{g}}\,, \textrm{\small{IC}}\,\}_{i} +  \{\,A_{\beta}\,, \beta\,\}_{\textrm{\ssz{R,B}}} \quad ,
 \end{array}
	\label{eq:nla_params}
\end{equation}
where we fit a galaxy bias and `integral constraint' ($\textrm{\small{IC}}$) to the galaxy clustering measured in each sample $i$. The integral constraint is a free parameter, taking the form of a small additive scalar applied to the clustering correlation function, to correct for the effects of a partial-sky survey window \citep{Roche1998}. Subscripts $\textrm{\small{R, B}}$ denote a red and blue version of each parameter, which are fit to all relevant samples. This brings the total number of parameters to 14 (16) for the NLA (NLA-$\beta$)\footnote{6 colour and redshift samples $i$ gives 12 clustering parameters (galaxy biases and integral constraints), plus a red and a blue amplitude for 14 in total. Another 2 luminosity scaling parameters makes 16.}. Previous dedicated IA studies have used galaxy clustering to fit and fix galaxy bias (e.g. \citealt{Joachimi2011}) -- we instead opt to marginalise over galaxy biases and integral constraints, thereby propagating our uncertainty in these parameters into our IA model constraints. Since our samples form independent datasets, by virtue of colour separation and disjoint areas, we can reduce the dimensionality of the problem by fitting our models to red and blue samples separately.

We choose not to include a redshift power-law scaling $\eta_{\textrm{other}}$ in our models, as has been done in previous works (\citealt{Hirata2007}, \citealt{Joachimi2011}, \citealt{Mandelbaum2011}), since the redshift baseline of our measurements is short -- GAMA starts to become sparse after $z\sim{0.4}$. While the results of previous work do not preclude the possibility of a significant redshift evolution, we argue that there is good reason to expect it to be small. Tidal torque theories suggest angular momentum generation as the source of spiral galaxy alignments. Since the spinning-up of a proto-galaxy halo is a perturbative effect, these alignments exist in the initial conditions of the matter field. After collapse of the overdense region, the angular momentum of the galaxy dominates over tidal torquing effects, and the galaxy orientation should be `frozen-in'. Subsequently, only merger events should change the orientation of the galaxy.

Mergers would be expected to erase the memory of previous alignments, disrupt galaxy and halo angular momenta and prompt a relaxation phase. The system should relax into a configuration with a reduced spin magnitude, diluting the quadratic alignment signature \citep{Cervantes-Sodi2010}. However, with merger timescales much shorter than relaxation, the spiral quickly transitions to a pressure-supported elliptical. The stellar distribution will then gradually re-equilibrate according to the ellipsoidal halo potential, itself moulded by the tidal field.

Therefore we might expect to observe `fixed' blue galaxy alignments, opposite red galaxy alignments with their evolution tied to the tidal field (and divided out of our models by the growth factor), or some diluted middling alignment for transitioning galaxies, where the change of sign takes the amplitude close to zero. \cite{Joachimi2011} constrain $\eta_{\textrm{other}}$ to be consistent with zero for early-type galaxies over a long redshift baseline. \cite{Mandelbaum2011} analysed late-type galaxy alignments in the WiggleZ survey \citep{Drinkwater2010}, with SDSS shapes, and also found $\eta_{\textrm{other}}$ to be consistent with zero. Furthermore, their null detection at a mean redshift $\bar{z}\sim0.6$ was recently matched by a null detection from the FastSound galaxy redshift survey \citep{Tonegawa2015} at $z\sim{1.4}$ \citep{Tonegawa2017}, suggesting no strong evolution of spiral galaxy alignments. Considering all of the above, we suggest that a physically motivated prior on $\eta_{\textrm{other}}$ should be narrow and centred on zero.

\section{IA CONSTRAINTS FOR FLUX-LIMITED SAMPLES}
\label{sec:results}

\begin{figure}
	\includegraphics[width=\columnwidth]{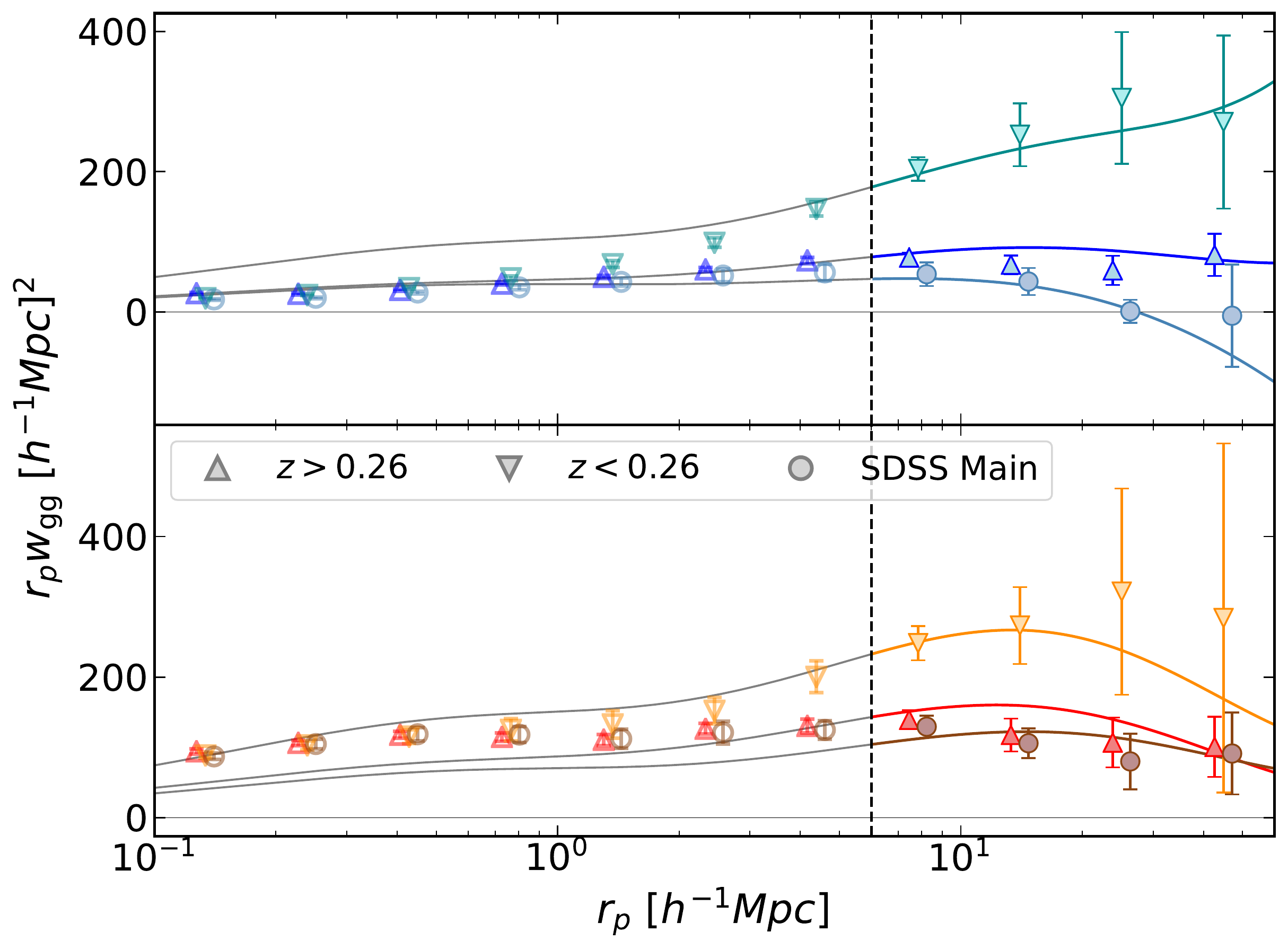}
    \caption{Measured galaxy clustering for our blue (\emph{top}) and red (\emph{bottom}) galaxy samples. Solid curves illustrate the best-fit linear clustering per sample (Eq. \ref{eq:gg_hankel}). The vertical dashed line indicates $r_{p}=6\mpc$, below which scales are excluded from fitting (Section \ref{sec:likelihoods}).}
    \label{fig:clus_measurements}
\end{figure}

\begin{figure}
	\includegraphics[width=\columnwidth]{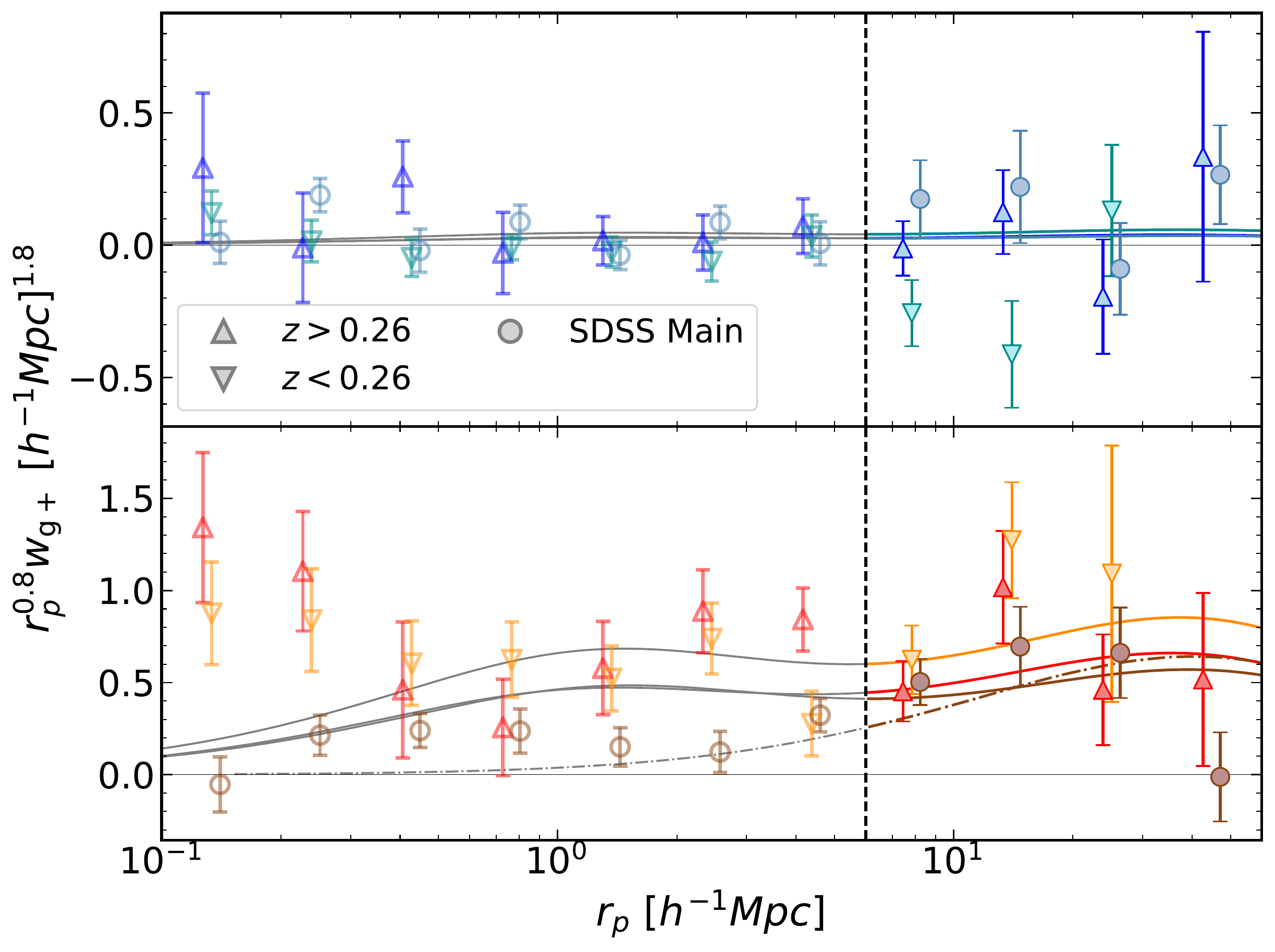}
    \caption{Measured galaxy position-intrinsic shear correlations for our blue (\emph{top}) and red (\emph{bottom}) galaxy samples. Best-fit NLA models are shown as solid curves, and the vertical dashed line indicates $r_{p}=6\mpc$, below which scales are excluded from fitting (Section \ref{sec:likelihoods}). The best-fit LA model to SR is shown as a dot-dashed line.}
    \label{fig:ia_measurements}
\end{figure}

\begin{figure}
	\includegraphics[width=\columnwidth]{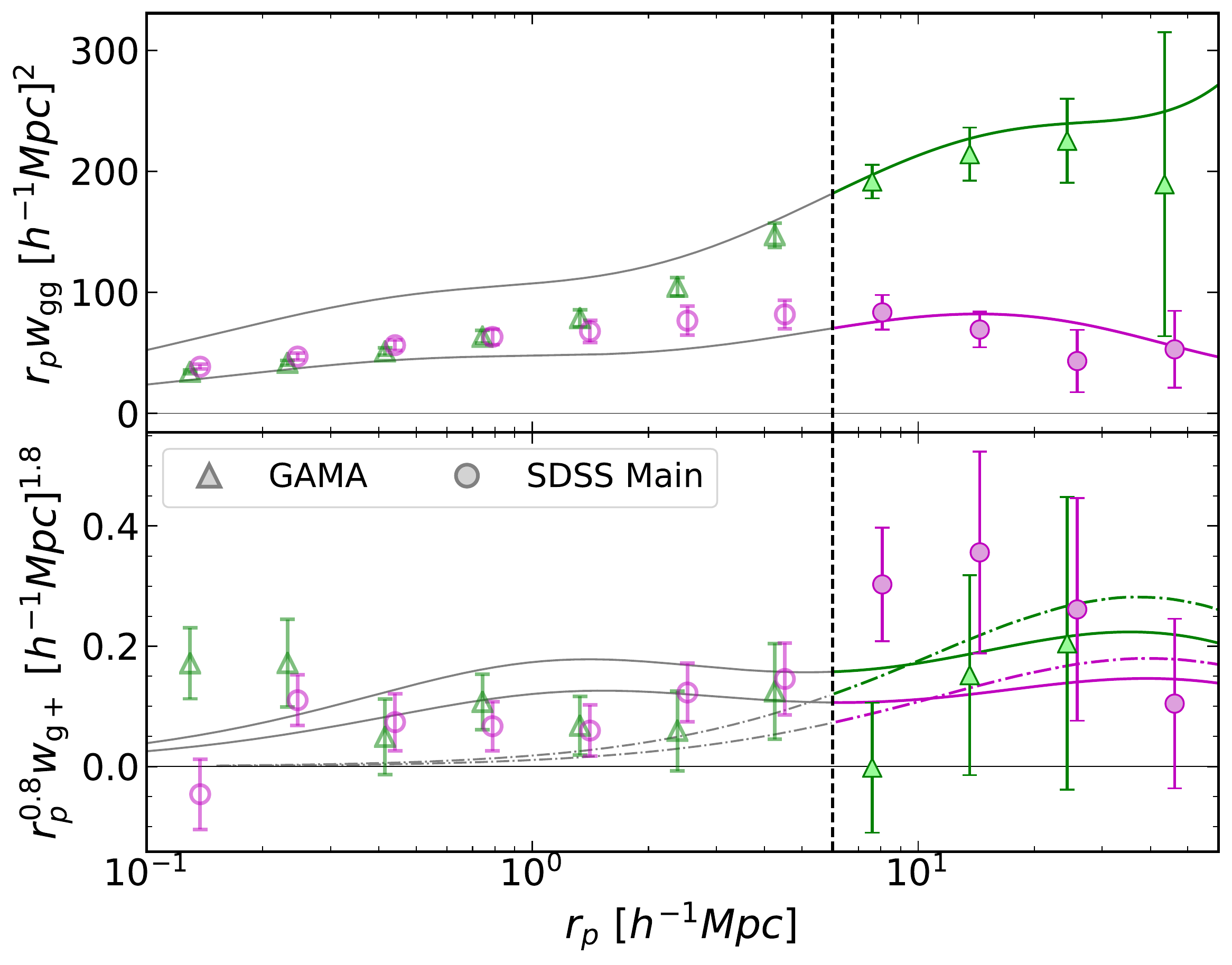}
    \caption{Galaxy clustering (\emph{top}) and position-intrinsic shear correlations (\emph{bottom}) measured in the full KiDS$+$GAMA and SDSS Main datasets. Solid lines illustrate the best-fit NLA model, and dot-dashed lines the LA. The vertical dashed line indicates $r_{p}=6\mpc$, below which scales are excluded from fitting (Section \ref{sec:likelihoods}).}
    \label{fig:hh_plot}
\end{figure}

\begin{figure*}
	\includegraphics[width=\textwidth]{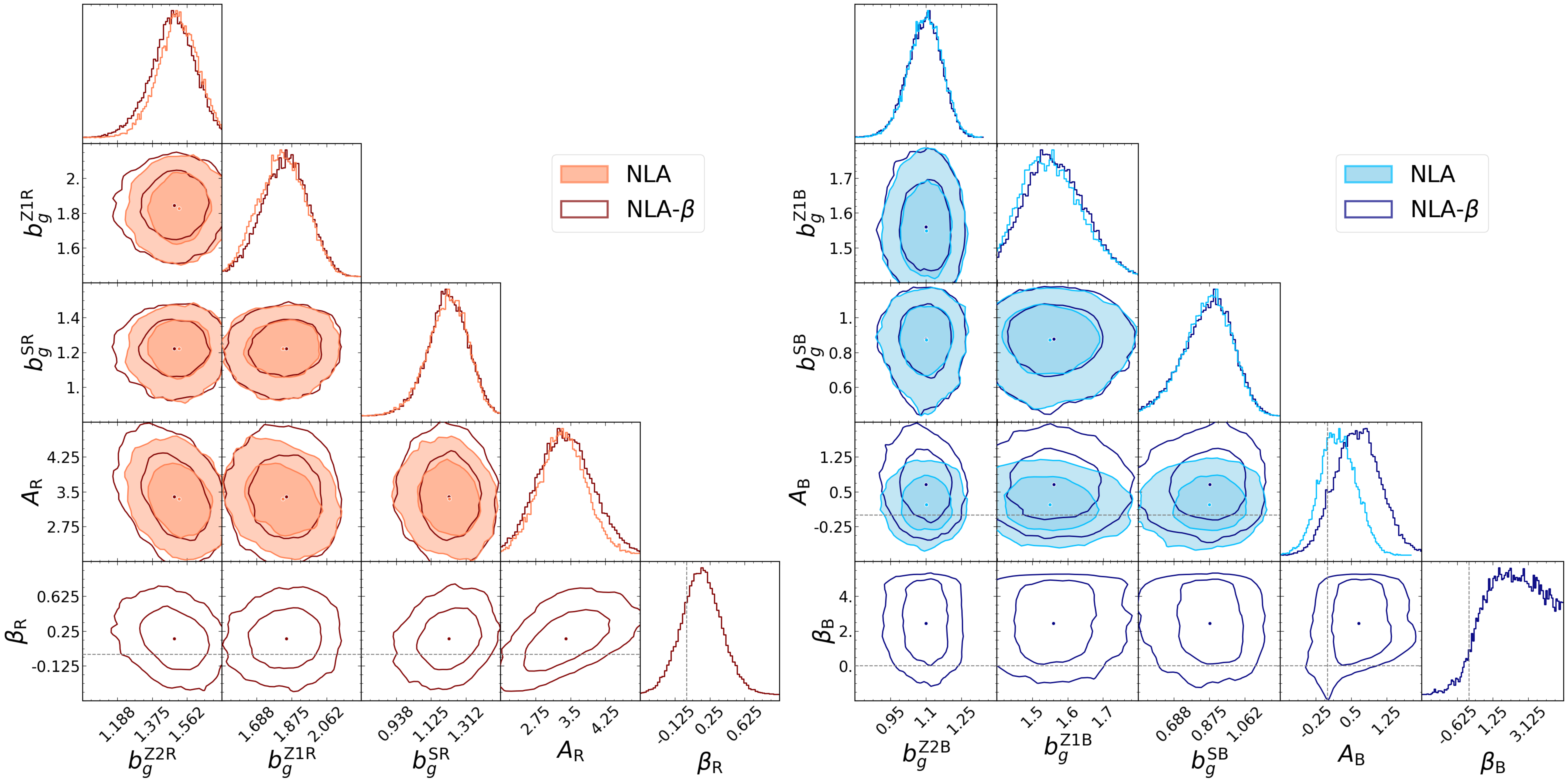}
    \caption{Posterior probability contours of our fitted galaxy bias $b_{\rm{g}}$, NLA amplitude $A$ and luminosity power-law $\beta$ parameters, for red (\emph{left}) and blue (\emph{right}) galaxies. The filled (unfilled) contours are for the NLA (NLA-$\beta$) models. Dashed grey lines mark values of zero for IA parameters.}
    \label{fig:ia_contours}
\end{figure*}

With our aim to motivate tighter, more realistic priors for intrinsic alignment parameters, we fit the standard and the luminosity-dependent N/LA models to galaxy position-intrinsic shear and clustering correlations in KiDS$+$GAMA and SDSS Main. We compute signal detection significances across all scales, and restrict fits of the models to transverse scales $>6\mpc$. Our various measurements are shown in Figures \ref{fig:clus_measurements}, \ref{fig:ia_measurements}, \ref{fig:hh_plot}. The results of fitting are shown in Figures \ref{fig:ia_contours} \& \ref{fig:indiv_aia_plot} and Table \ref{tab:constraints}.

\begin{center}
\begin{table*}
	\small
	\caption{NLA model parameter and galaxy bias $1\rm{D}$ marginalised constraints for our samples, with $68\%$ confidence intervals and the reduced $\chi^{2}$ ($\chi_{\nu}^{2}=\chi^{2}$ per degree of freedom) statistics for the global fit. $A_{\beta}$ denotes the alignment amplitude parameter of the NLA-$\beta$ model (Eq. \ref{eq:nla-beta}). The mean galaxy biases shift slightly with the NLA-$\beta$ -- these changes are insignificant within statistical errors on these parameters, and are not shown in the table. Bracketed numbers indicate properties of density samples, as opposed to shapes samples. `G' and `S' denote GAMA and SDSS samples, respectively.}
	\label{tab:constraints}
	\def\arraystretch{1.5}
	\begin{tabular*}{\textwidth}{lccclcccccc} 
		\hline
		 Sample & $\langle{}z\rangle$ & $\langle{}L/L_{\rm{piv}}\rangle$ & $b_{\rm{g}}$ & $\quad \quad \Aia$ & $\chi^{2}_{\nu}$ & $p(>\chi^{2})$ & $A_{\beta}$ & $\beta$ & $\chi^{2}_{\nu}$ & $p(>\chi^{2})$ \\
		\hline
		\hline
GAMA full & 0.23 (0.24) & 0.51 (0.70) & $1.57_{-0.09}^{+0.08}$ & \rdelim\}{2}{1mm} \,\, \multirow{2}{*}{$1.06_{-0.46}^{+0.47}$} & \multirow{2}{*}{1.32} & \multirow{2}{*}{0.21} & \multirow{2}{*}{$0.87_{-1.43}^{+4.00}$} & \multirow{2}{*}{$2.06_{-2.82}^{+2.20}$} & \multirow{2}{*}{1.12} & \multirow{2}{*}{0.34} \\
SDSS Main full & 0.11 (0.11) & 0.22 (0.22) & $0.94_{-0.11}^{+0.10}$ &   &   &   &   &   &   &   \\
G: $z>0.26$, blue & 0.33 (0.33) & 1.06 (1.09) & $1.10_{-0.07}^{+0.07}$ & \rdelim\}{3}{1mm} &   &   &   &   &   &   \\
G: $z<0.26$, blue & 0.15 (0.17) & 0.21 (0.36) & $1.55_{-0.08}^{+0.09}$ & $\quad 0.21_{-0.36}^{+0.37}$ & 1.37 & 0.14 & $0.65_{-0.51}^{+0.50}$ & $2.47_{-1.59}^{+1.68}$ & 1.34 & 0.17 \\
S: blue & 0.09 (0.09) & 0.14 (0.14) & $0.88_{-0.14}^{+0.12}$ &   &   &   &   &   &   &   \\
G: $z>0.26$, red & 0.33 (0.33) & 1.47 (1.48) & $1.52_{-0.11}^{+0.11}$ & \rdelim\}{3}{1mm} &   &   &   &   &   &   \\
G: $z<0.26$, red & 0.17 (0.18) & 0.50 (0.56) & $1.84_{-0.12}^{+0.12}$ & $\quad 3.18_{-0.45}^{+0.46}$ & 1.28 & 0.20 & $3.40_{-0.56}^{+0.59}$ & $0.18_{-0.22}^{+0.20}$ & 1.34 & 0.17 \\
S: red & 0.12 (0.12) & 0.29 (0.29) & $1.19_{-0.11}^{+0.11}$ &   &   &   &   &   &   &   \\
		\hline
	\end{tabular*}
\end{table*}
\end{center}

\subsection{Clustering}

Relating the matter-intrinsic power spectrum $P_{\delta\rm{I}}$ to \wgp requires estimations of the galaxy bias $b_{\rm{g}}$ of our density tracers. Hence we measure galaxy clustering in our density samples and perform fits of a linear, scale-independent bias with the full matter power spectrum (Eq. \ref{eq:gg_hankel}). We verify that our clustering pipeline reproduces the GAMA measurements of \cite{Farrow2015} for their sample selection.

Figure \ref{fig:clus_measurements} shows our measurements of $w_{\rm{gg}}$ in GAMA and SDSS, with best-fit linear clustering overlaid. Our fits include the integral constraint correction (Section \ref{sec:likelihoods}), which is small ($|\textrm{\small{IC}}| \lesssim 3\mpc$) and therefore negligible on small scales. 
Fits of the linear clustering model are restricted to scales $>6\mpc$, indicated by vertical dashed lines. Our sample galaxy bias fits are summarised in Table \ref{tab:constraints}. The biases form a consistent and expected picture -- more luminous samples are more biased at the same redshifts.

\subsection{Alignments}

Figure \ref{fig:ia_measurements} shows our colour-split measurements of $w_{\rm{g+}}$, overlaid with the best-fitting NLA (solid lines). We also perform fits to our data with the LA model, shown as dot-dashed lines in Figure \ref{fig:ia_measurements} (to SR only) and Figure \ref{fig:hh_plot}. Table \ref{tab:null_tests} lists signal detection significances for the alignment signals and systematics tests (described in Section \ref{sec:estimators}).

\begin{table}
 \tiny
 \begin{center}
 	\caption{Reduced $\chi^{2}$ statistics to assess the significance of signal detections against the null hypothesis (i.e. a zero-signal), for \wgp and for systematics tests; \wgx and \wgp limited to large line-of-sight separations ($60\leqslant|\Pi|\leqslant90\mpc$), denoted $\Pi+$. Bracketed numbers indicate the statistics when restricting to the \rp -scales $>6\mpc$ which are fitted in the analysis.}
 	\label{tab:null_tests}
 	\def\arraystretch{1.4}
 	\begin{tabular}{lcccc} 
 		\hline
 		 Sample & Signal & $\chi^{2}_{\nu\,,\textrm{\ssz{null}}}$ & $p(>\chi^{2})$ & $\sigma$\\
 		\hline
        \hline
Blue total & $w_{\rm{g}+}$ & 0.85 (1.24) & 0.71 (0.25) & 0.37 (1.14) \\
\hline
GAMA, & $w_{\rm{g}+}$ & 0.31 (0.38) & 0.98 (0.82) & 0.02 (0.22) \\
$z>0.26$, & $w_{\rm{g}+} (\Pi+)$ & 0.20 (0.00) & 0.98 (1.00) & 0.03 (0.00) \\
blue & $w_{\rm{g}\times}$ & 0.93 (1.66) & 0.51 (0.16) & 0.66 (1.42) \\
\hline
GAMA, & $w_{\rm{g}+}$ & 0.85 (2.55) & 0.58 (0.05) & 0.56 (1.93) \\
$z<0.26$, & $w_{\rm{g}+} (\Pi+)$ & 0.41 (1.27) & 0.87 (0.28) & 0.16 (1.08) \\
blue & $w_{\rm{g}\times}$ & 0.44 (0.22) & 0.94 (0.93) & 0.08 (0.09) \\
\hline
SDSS Main, & $w_{\rm{g}+}$ & 1.38 (1.12) & 0.17 (0.34) & 1.36 (0.95) \\
blue & $w_{\rm{g}+} (\Pi+)$ & 0.44 (0.78) & 0.85 (0.46) & 0.19 (0.74) \\
  & $w_{\rm{g}\times}$ & 1.14 (1.30) & 0.33 (0.27) & 0.98 (1.11) \\
\hline
Red total & $w_{\rm{g}+}$ & 5.03 (6.86) & 0.00 (0.00) & 8.93 (6.79) \\
\hline
GAMA, & $w_{\rm{g}+}$ & 4.03 (4.37) & 0.00 (0.00) & 4.51 (3.17) \\
$z>0.26$, & $w_{\rm{g}+} (\Pi+)$ & 0.74 (2.97) & 0.62 (0.05) & 0.50 (1.95) \\
red & $w_{\rm{g}\times}$ & 0.31 (0.42) & 0.98 (0.79) & 0.02 (0.26) \\
\hline
GAMA, & $w_{\rm{g}+}$ & 6.27 (8.85) & 0.00 (0.00) & 6.09 (4.48) \\
$z<0.26$, & $w_{\rm{g}+} (\Pi+)$ & 0.32 (0.30) & 0.93 (0.74) & 0.09 (0.33) \\
red & $w_{\rm{g}\times}$ & 0.75 (1.20) & 0.69 (0.31) & 0.40 (1.02) \\
\hline
SDSS Main, & $w_{\rm{g}+}$ & 4.90 (7.86) & 0.00 (0.00) & 5.29 (4.71) \\
red & $w_{\rm{g}+} (\Pi+)$ & 0.84 (0.87) & 0.54 (0.42) & 0.61 (0.81) \\
  & $w_{\rm{g}\times}$ & 0.28 (0.20) & 0.99 (0.94) & 0.01 (0.08) \\
 		\hline
 	\end{tabular}
 \end{center}
\end{table}

\subsubsection{Signals \& NLA results}
\label{sec:nla_results}

We find blue galaxy alignments to be consistent with zero, in agreement with previous studies of this population (\citealt{Mandelbaum2011}, \citealt{Tonegawa2017}). The NLA-$\beta$ amplitude $A_{\beta}$ and power-law $\beta$ are also consistent with zero, at $95\%$ confidence. For blue galaxies on the whole, or for individual blue samples, we make no significant detections of \wgp, whether restricting to linear scales, or considering the full range in $r_{p}$ (Table \ref{tab:null_tests}).
Fits to GAMA-only: $\Aia = 0.04^{+0.44}_{-0.42}$ , and SDSS-only: $\Aia = 1.03^{+0.90}_{-0.85}$, are consistent with each other, and the total-fit, at $68\%$ confidence.

In agreement with previous work (\citealt{Hirata2007}, \citealt{Joachimi2011}), we measure a significantly positive amplitude of alignments for red galaxies, in both modes of fitting and at $>95\%$ confidence. The total significance of detection we find for red galaxy alignments is close to $9\sigma$ over the full range in $r_{p}$, and $6.79\sigma$ when limited to linear scales ($\geqslant6\mpc$). The $\beta$ luminosity-scaling is found to be comfortably consistent with zero, and thus results in a poorer fit (owing to a lost degree of freedom) than for the 1-parameter NLA. This is in contrast with previous observations of near-linear scalings of red galaxy/LRG alignments with luminosity (e.g. \citealt{Hirata2007}, \citealt{Joachimi2011}, \citealt{Singh2015}). The perturbative IA model of \cite*{Blazek2015} uncovered additional contributions to the observed large-scale intrinsic shape correlation, arising from source density weighting \citep{Hirata2004a} -- as galaxies preferentially exist in overdense space, our sampling of the intrinsic ellipticity field is necessarily biased, as briefly discussed in Section \ref{sec:likelihoods}. This contribution was found to be galaxy bias-dependent, and mooted as responsible for such observed luminosity-scalings -- indeed we measure SDSS red to have the weakest alignment signature (see Section \ref{sec:indivAia}) of our red samples, although the significance of this is questionable. A GAMA-only fit results in a slightly higher red galaxy alignment amplitude of $\Aia = 3.52^{+0.60}_{-0.56}$, whilst SDSS-only returns $\Aia = 2.50^{+0.77}_{-0.73}$, again comfortably consistent with each other and the total fit.

For the `full' (all-galaxy) samples, we measure a positive NLA amplitude at just over $95\%$ confidence, whilst the NLA-$\beta$ is poorly constrained, owing to a sparse luminosity baseline. A point of interest is the apparently larger amplitude of \wgp measured for SDSS, compared with GAMA, for which the N/LA models are unable to account -- the green and purple curves in Figure \ref{fig:hh_plot} differ only by their dependence on the (subdominant) weight function $\mathcal{W}(z)$ and the fitted galaxy bias per sample. Individual fits to these samples yield $\Aia = 0.26^{+0.63}_{-0.62}$, and $\Aia = 2.01^{+0.79}_{-0.71}$, for GAMA and SDSS, respectively -- mildly discrepant at $\sim1.8\sigma$. GAMA is brighter, and constrained to be more biased, than SDSS, seemingly ruling-out luminosity/bias-dependences as explanations. It must, however, be noted that these all-galaxy signals constitute muddy combinations of clearly dichotomous alignment signatures, and that GAMA and SDSS sample different environments -- something we explore in the next section.

A primary motive for this work was to take advantage of highly complete, flux-limited data in order to constrain IA as it pertains to cosmic shear contamination. The only comparable analyses to date are the SDSS Main studies of \cite{Mandelbaum2006} -- M06, and \cite{Hirata2007}\footnote{\cite{Hirata2007} also studied LRGs - we only discuss their work on the flux-limited SDSS Main sample.} -- H07, each of which was conducted slightly differently to this work. For example, neither study made use of the N/LA models as they are typically formulated today, allowing for no easy comparison of fitted alignment amplitudes $\Aia$. In any case, our sample selections are also quite different -- both M06 \& H07 made use of the long luminosity baseline in SDSS to create subsamples, and whilst H07 also split their samples into red/blue galaxies, their cut was performed using observer-frame magnitudes. Nevertheless, we make some broadly similar findings; H07 made robust detections of IA in red galaxies, as did M06 for their brightest sample, itself dominated by red galaxies. Additionally, H07 also failed to make a significant detection for blue galaxies.

We do however seem to find some indirect disagreement in the alignment amplitude vs. sample luminosity trend inferred from the data. Each of M06 \& H07 saw trends of increasing signal amplitudes with sample luminosity, whilst we find no evidence for luminosity evolution in our model fits. Furthermore, the far brighter Z2R sample exhibits an amplitude of alignment entirely consistent with the Z1R fit, and as mentioned above, we measure a larger amplitude of alignment for the fainter (uncut) SDSS sample than for GAMA. We explore these individual fits, and how they correlate with sample properties, in Section \ref{sec:indivAia}.

\subsubsection{Individual sample fits}
\label{sec:indivAia}

\begin{figure*}
	\includegraphics[width=\textwidth]{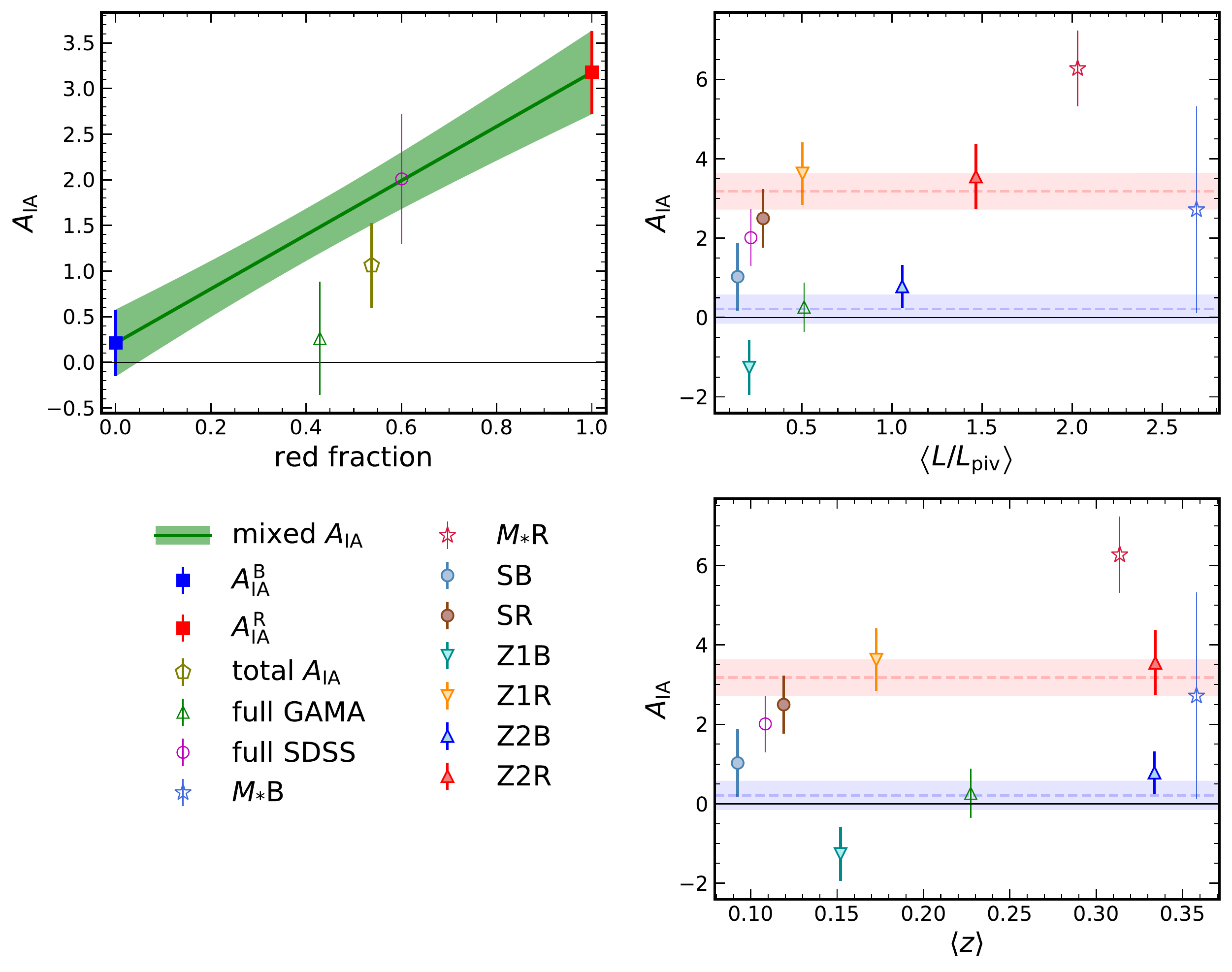}
    \caption{Constraints on the NLA model alignment amplitude $\Aia$, from various subsamples of GAMA and SDSS (Table \ref{tab:sample_details}), plotted against sample properties. The constraints illustrated here are also given in Table \ref{tab:indiv_constraints}}. \emph{Top left:} $\Aia$ vs. shape sample red galaxy fraction. We interpolate (green line/shading) between our fits to blue (blue square) and red (red square) galaxy samples, according to $\Aia=A^{\rm{R}}_{\textrm{\ssz{IA}}}f_{\rm{red}} + A^{\rm{B}}_{\textrm{\ssz{IA}}}(1-f_{\rm{red}})$, where $f_{\rm{red}}$ is the red fraction and we assume linearity in the contributions of galaxy populations to the total alignment signal/amplitude. The inconsistency of mixed-sample signals (open points) with this interpolation is due to variable contributions of satellite galaxies -- this is discussed in Section \ref{sec:indivAia}. \emph{Top right:} $\Aia$ vs. shape sample luminosity (as a ratio to the pivot $L_{\rm{piv}}\sim\num{4.6e10}L_{\sun}$, corresponding to absolute $r$-band magnitude $M_{r}=-22$). \emph{Bottom right:} $\Aia$ vs. shape sample mean redshift. All plotted data points illustrate the mean and $68\%$ confidence interval of $1\rm{D}$ marginalised posterior distributions on $\Aia$, after fitting to relevant alignment/clustering signals. Only the filled points are independent of each other; each of the open points is in some way correlated with the others. Dashed lines and shading indicate the mean and $68\%$ CI of the total-colour fits, highlighting the type-dependence of alignments.
    \label{fig:indiv_aia_plot}
\end{figure*}

We make additional, individual fits of $\Aia$ to each of our galaxy samples, to gain further insight into trends with colour, luminosity and redshift. Figure \ref{fig:indiv_aia_plot} illustrates the results of fitting individual amplitudes to (i, squares in top left panel) red and blue signals, (ii, filled points in right panels) signals from each of our colour/redshift-split samples in GAMA and SDSS, (iii, unfilled triangles/circles) individual signals from uncut GAMA and SDSS, (iv, pentagon in top left panel) all signals from the uncut samples, and (v, stars in right panels) signals from GAMA galaxies with stellar-mass $M_{*}\geqslant10^{11}M_{\sun}$. Only the filled data points are independent of each other, as the unfilled points are each fitted to some collection/subset of the independent samples -- Table \ref{tab:indiv_constraints} details the constraints from each sample, with independent samples denoted by $\dagger$.

In each panel of the figure, there is a clear dichotomy in the fitted amplitudes for red and blue galaxies, highlighted in the right-hand panels by dashed lines and shading. The top right panel shows $\Aia$ vs. sample luminosity, and reveals a vaguely positive correlation in the filled data points, but at very low significance, especially if one (i) considers blue and red separately, and (ii) notes that the Z1B fitted amplitude is anomalously low with respect to the other blue sample amplitudes\footnote{We note that the Z1B amplitude is consistent with zero at $95\%$ confidence, and that this signal (downward cyan triangles in the top panel of Fig. \ref{fig:ia_measurements}) is not found to be a particularly significant detection, at $<2\sigma$ (Table \ref{tab:null_tests}). Additionally, the signal becomes comfortably consistent with zero upon removal of the faint-limit we apply to our GAMA density samples (explained in Appendix \ref{sec:app_mice_errors}), which affects the Z1B sample far more significantly than each of the others combined. This could be interpreted as follows; the Z1B sample shows a net tangential alignment at $\sim1.7\sigma$, but \emph{only} when excluding the faintest ($\sim27\%$ here) galaxies from the density sample. However, the faint-limit is part of our clustering covariance estimation (see Appendix \ref{sec:app_mice_errors}) -- removing it may invalidate the clustering fits which anchor the galaxy bias, so this interpretation must be taken with moderation. A linear-scale tangential alignment of blue galaxies, dependent on the bias of the density tracer, is an interesting result which would call for further work. However, it should be noted that (i) tidal torquing mechanisms ought to be weak on these scales, so this signal is not expected, (ii) the significance of the negative amplitude is low, and (iii) the signal itself lacks a clear detection.}.

The bottom right panel shows $\Aia$ vs. sample mean-redshifts, with any correlation even less pronounced. To date, no direct IA analyses have found evidence for redshift evolution of intrinsic alignments (\citealt{Joachimi2011}, \citealt{Mandelbaum2011}, \citealt{Tonegawa2017}), and our results seem to agree -- although it should be noted that our baseline is short, and limited to the relatively near universe. Some works have reported evidence for scaling of IA with sample luminosity (\citealt{Hirata2007}, \citealt{Joachimi2011}, \citealt{Singh2015}), findings unsupported by our measurements -- we do make a clean detection for massive, red GAMA galaxies (red stars), at $9.1\sigma$ and with a large fitted amplitude of alignment, but these galaxies are effectively a subset of (primarily) the Z2R sample. Thus the large-$M_{*}$ points are highly correlated with their high-redshift counterparts; these points (upward triangles) disagree with the notion of luminosity dependence. As discussed above, it may be that such an observed dependence is due to environmental properties which correlate with luminosity. Our data points might weakly support this assertion for red galaxies, given that we constrain SDSS red to be less biased than the red GAMA samples (see Table \ref{tab:constraints}), however the significance is extremely low; more work is needed for a concrete answer to this question.

In the top left panel, we interpolate between the fitted red and blue alignment amplitudes according to

\begin{equation}
	\Aia = A^{\rm{R}}_{\textrm{\ssz{IA}}}f_{\rm{red}} + A^{\rm{B}}_{\textrm{\ssz{IA}}}(1-f_{\rm{red}})
    \label{eq:redfrac}
\end{equation}
where $f_{\rm{red}} \in [\, 0\,, 1\,]$ is the sample red fraction, and we assume that the red and blue galaxy populations contribute linearly to the measurable alignment of the full sample. Thus we provide predictions\footnote{Inserting our IA model constraints from Tables \ref{tab:constraints} or \ref{tab:la_constraints} into the functional form of Eq. \ref{eq:redfrac}, one can derive an expected confidence interval on $A_{\rm{IA}}$, for the NLA or LA, given a sample red-fraction.} for the $\Aia$ one might measure in a flux-limited sample of mixed galaxy-type, given the red galaxy fraction, and \emph{provided} that the red/blue dichotomy is the dominant driver of the alignment profile.

\begin{figure*}
	\includegraphics[width=\textwidth]{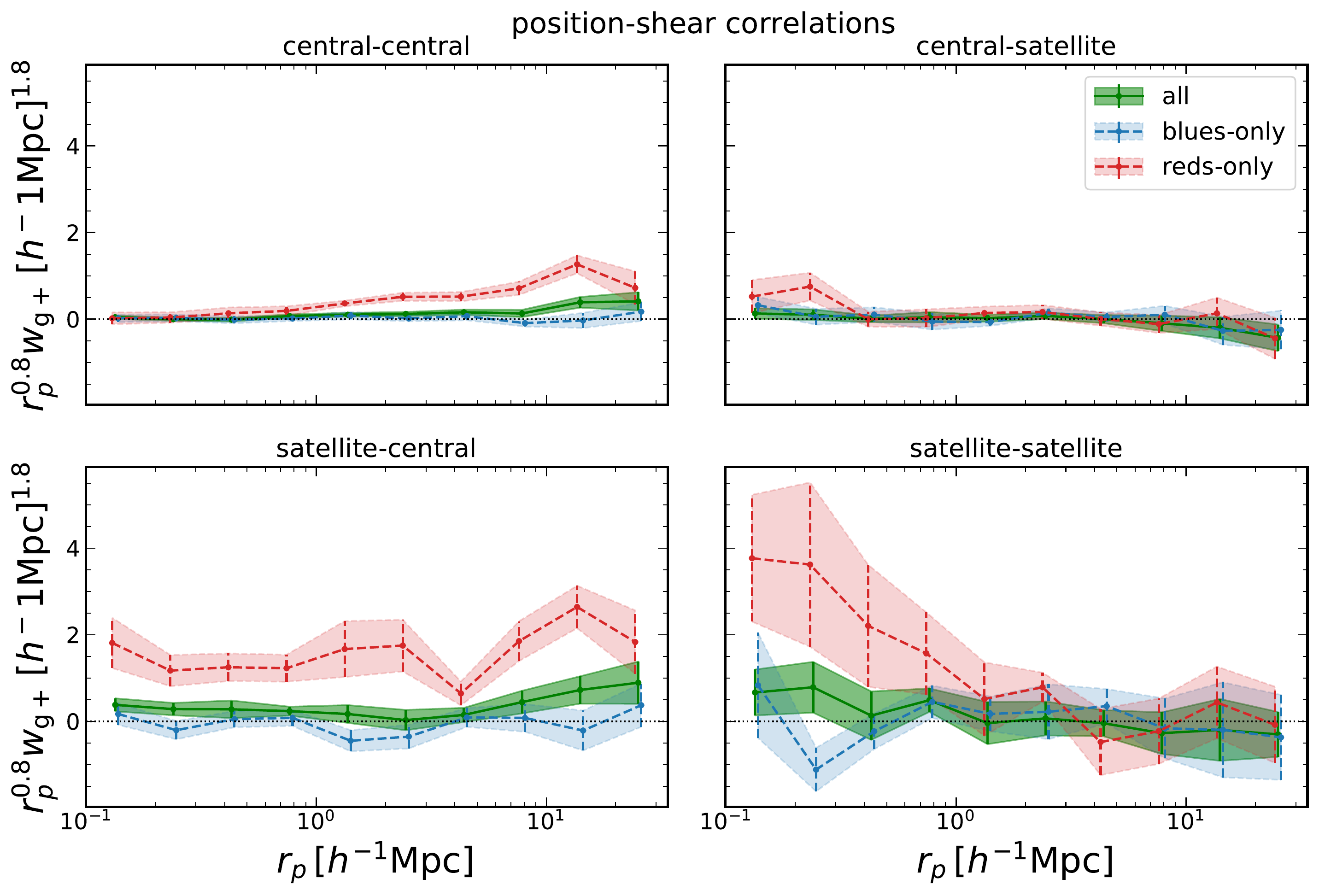}
	\caption{Various position-intrinsic shear correlations measured between GAMA samples of exclusively central or satellite galaxies, with errors estimated via jackknife. The title of each panel indicates the central/satellite composition of the position-shear (i.e. density-shape) samples, and we measure correlations in the mixed samples, and for red- (red dashed) and blue-only (blue dashed) subsets. }
    \label{fig:censat_plot}
\end{figure*}

Shown also in the top-left panel are the two $\Aia$ fitted to the total GAMA (green triangle) and total SDSS (pink circle) signals (shown in the bottom panel of Figure \ref{fig:hh_plot}), and the single amplitude fitted to both signals (gold pentagon). One clearly sees that GAMA galaxies are less radially aligned than is predicted by the interpolation. We find this discrepancy to be driven by a significant fraction of satellite galaxies in GAMA, with differing alignment behaviour -- previous work has found satellite galaxy alignments to be weaker than those of central galaxies, or altogether non-existent (\citealt{Sifon2015}, \citealt{Singh2015}, \citealt{Huang2018}), in particular when considering larger pair-separations as we do when fitting our models.



Figure \ref{fig:censat_plot} breaks down the central\footnote{We count field galaxies as centrals, assuming that their satellites are simply too faint to be detected.} and satellite, red and blue contributions to the total GAMA alignment signature. The low amplitude fitted to GAMA is simply understood in this context -- the left-hand panels demonstrate linear-scale ($\gtrsim$ a few $\mpc$) alignments to be sourced entirely by red central galaxy \emph{shapes}. All other galaxy shapes -- red satellites, blue centrals and blue satellites -- are unaligned on these scales (seen in all panels), and thus dampen the overall alignment signature with zero-mean white noise. Thus the linear-scale alignment correlation can be thought of as `set' by the red central galaxies, and then repeatedly damped upon the inclusion of other species; blue galaxies have zero-signal, and so force an effective rescaling of \wgp by a factor $\sim{f_{{\rm{red}}}}$; red satellites do source a strong signal (bottom-left panel) through the inclusion of their positions (which are highly correlated with centrals on these scales), but this is slashed by their own lack of alignment (right-panels), making the overall dampening a more complicated function of red central/satellite fractions.

Being $\sim2$ magnitudes shallower than GAMA, and at less than a tenth of the on-sky density, SDSS is comparatively deficient in fainter satellite galaxies at low redshift (see right-panel of Figure \ref{fig:sample_spaces}). Thus the linear-scale alignment dampening described above is more severe for GAMA than for SDSS, explaining the behaviour seen in Figures \ref{fig:hh_plot} \& \ref{fig:indiv_aia_plot}. Indeed, we find an alignment amplitude fit to the mixed central galaxy signal in GAMA (Figure \ref{fig:censat_plot}, top-left panel, green curve) to sit comfortably atop the interpolation of Figure \ref{fig:indiv_aia_plot}, with an almost unchanged red fraction.

Inspecting the signals themselves, we note first that blue galaxies exhibit null signals under every division of the data. We also see that red central galaxies align radially with each other at large-$r_{p}$ (top-left), and with satellites at small-$r_{p}$ (bottom-left) -- we re-measure this signal with $|\Pi_{\rm{max}}| = 12\mpc$ to confirm that these centrals are aligning with \emph{their own} satellite distribution. In comparison, red satellites align strongly, but more noisily, with each other on smaller scales (bottom-right), and are elsewise unaligned. With these measurements we can make the following statements about red galaxies: satellite galaxies exist preferentially along the semi-major axis direction of the central galaxy, and the satellite galaxies are, on average, aligned with this direction. These are interesting considerations for future work, given that the satellite distribution is thought to trace that of the underlying dark matter.

Satellite considerations thus explain the discordance between the blue-to-red amplitude interpolation in Figure \ref{fig:indiv_aia_plot} and the amplitudes fitted to GAMA signals, and call for additional work; a motivated prior for the amplitude of intrinsic alignments in a cosmic shear study may need to consider not only the red fraction of the galaxy sample, but also the satellite fraction. Such population fractions will correlate with each other to an extent, with redshift as the universal galaxy population evolves, and with spatially variable limiting magnitudes for any given survey. To complicate matters further, \cite{Georgiou2018} find these influential red satellite galaxies to drive variation in measurable alignment signatures as a function of the passband of observation; cosmic shear studies in different bands can expect different contributions of alignments to shear signals. Thus predicting the IA contamination of shear in a galaxy survey is highly non-trivial.

\subsubsection{LA results}
\label{sec:la_results}

Linear alignment model fits to the data (fully detailed in Table \ref{tab:la_constraints}) result in consistency with analogous parameters from the NLA/NLA-$\beta$ at $68\%$ in all cases. The slightly larger amplitudes seen for red galaxies reflect the smaller amplitude of fluctuations in the linear matter power spectrum (see Section \ref{sec:nla}). This can be seen most clearly in the bottom panel of Figure \ref{fig:ia_measurements}, where the LA fit to SR (brown dot-dashed line) happens to closely match the NLA-Z1R fit (orange solid line) in amplitude. The linear model shows a clear deficit in power at scales $\lesssim20\mpc$, relative to the NLA. Consistency between the blue LA and NLA models is even stronger, as expected for null-signals.

The $\chi^{2}$ statistics in Tables \ref{tab:constraints} \& \ref{tab:la_constraints} purport the LA model to describe these data almost as well as the NLA on scales $\geqslant6\mpc$, though it is clear from the N/LA illustrations in Figures \ref{fig:ia_measurements} \& \ref{fig:hh_plot} that (i) neither model is sufficient to capture the complex variation of alignments as a function of galaxy sample properties and (ii) only the red GAMA signals would seem to explicitly prefer the enhancement offered by the NLA on scales of a few $\mpc$. The inclusion of blue galaxies efficiently washes out the \wgp signal on those scales (green points in Figure \ref{fig:hh_plot}), such that something in-between the N/LA models would appear closer to the truth. This result reaffirms the need for more complex modelling of IA in cosmic shear, highlighting the non-trivial contributions of various (i.e. colour, environment) sub-samples to overall alignment signatures.


\subsection{Systematics tests}

Table \ref{tab:null_tests} lists the detection significances of our measured signals -- \wgp, $w_{\rm{g+}}\{60 < |\Pi| < 90\mpc\}$ and \wgx -- across all $r_{p}$-scales, and when limited to the scales of fitting ($>6\mpc$; bracketed numbers). We make no significant ($>2\sigma$) detections of any systematic signals (see Section \ref{sec:estimators}) in our samples.


\section{IMPACT ON COSMOLOGY}
\label{sec:forecasting}

\begin{table}
\small
\begin{center}
	\caption{Gaussian priors on cosmological and IA/photo-$z$ nuisance parameters adopted for our Fisher forecasts. Centres are those of the fiducial cosmology, which includes the maximum likelihood points (not the $1\rm{D}$ marginals) of our IA analysis. The `Informative' set of priors for photo-$z$ bias parameters are approximations of constraints on equivalent parameters from the KiDS joint-probe analysis by \protect\cite{VanUitert2018}, with an extension for a fifth redshift bin.}
 	\label{tab:forecast_priors}
	\def\arraystretch{1.3}
	\begin{tabular}{lcc} 
		\hline
		Parameter & centre & width \\
		\hline
		\hline
		$h$ & 0.7 & 0.15 \\
		$n_{s}$ & 0.95 & 0.01 \\
		\hline
		$A^{\textrm{\ssz{B}}}_{\beta}$ & 0.58 & 20 \\
		$\beta_{\textrm{\ssz{B}}}$ & 3.15 & 5 \\
		\hline
		Modest: & & \\
		$a_{z_{1}}$ & 0 & 0.05 \\
		$a_{z_{2}}$ & 0 & 0.055 \\
		$a_{z_{3}}$ & 0 & 0.06 \\
		$a_{z_{4}}$ & 0 & 0.065 \\
		$a_{z_{5}}$ & 0 & 0.07 \\
		\hline
		Informative: & & \\
		$a_{z_{1}}$ & 0 & 0.036 \\
		$a_{z_{2}}$ & 0 & 0.042 \\
		$a_{z_{3}}$ & 0 & 0.048 \\
		$a_{z_{4}}$ & 0 & 0.054 \\
		$a_{z_{5}}$ & 0 & 0.062 \\
		\hline
	\end{tabular}
\end{center}
\end{table}

\begin{figure*}
	\includegraphics[width=\textwidth]{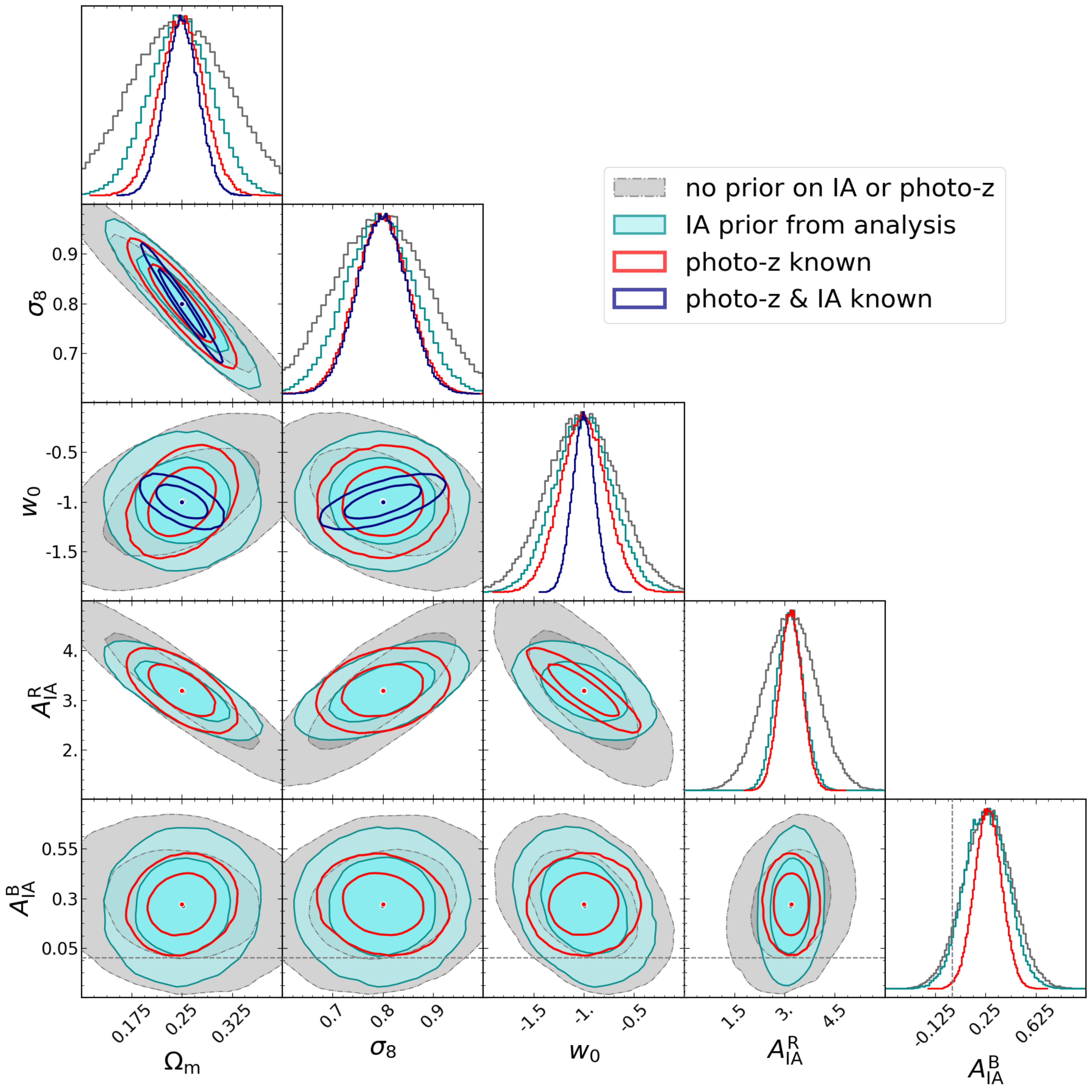}
    \caption{Fisher forecasted cosmological constraints for a KiDS-like survey, with (cyan) and without (grey) the application of our derived IA priors, assuming intrinsic alignments obey the non-linear alignment (NLA) model. Filled contours correspond to forecasts without any priors upon photo-z bias parameters (c.f. the `Modest' and `Informative' prior cases in Table \ref{tab:forecast_priors}). Forecasts with photo-$z$ bias fixed to zero are represented by navy and red unfilled contours, where navy also assumes perfect knowledge of IA model parameters. Dashed grey lines mark values of zero for nuisance parameters.}
    \label{fig:forecast_nobeta}
\end{figure*}

\begin{figure*}
	\includegraphics[width=\textwidth]{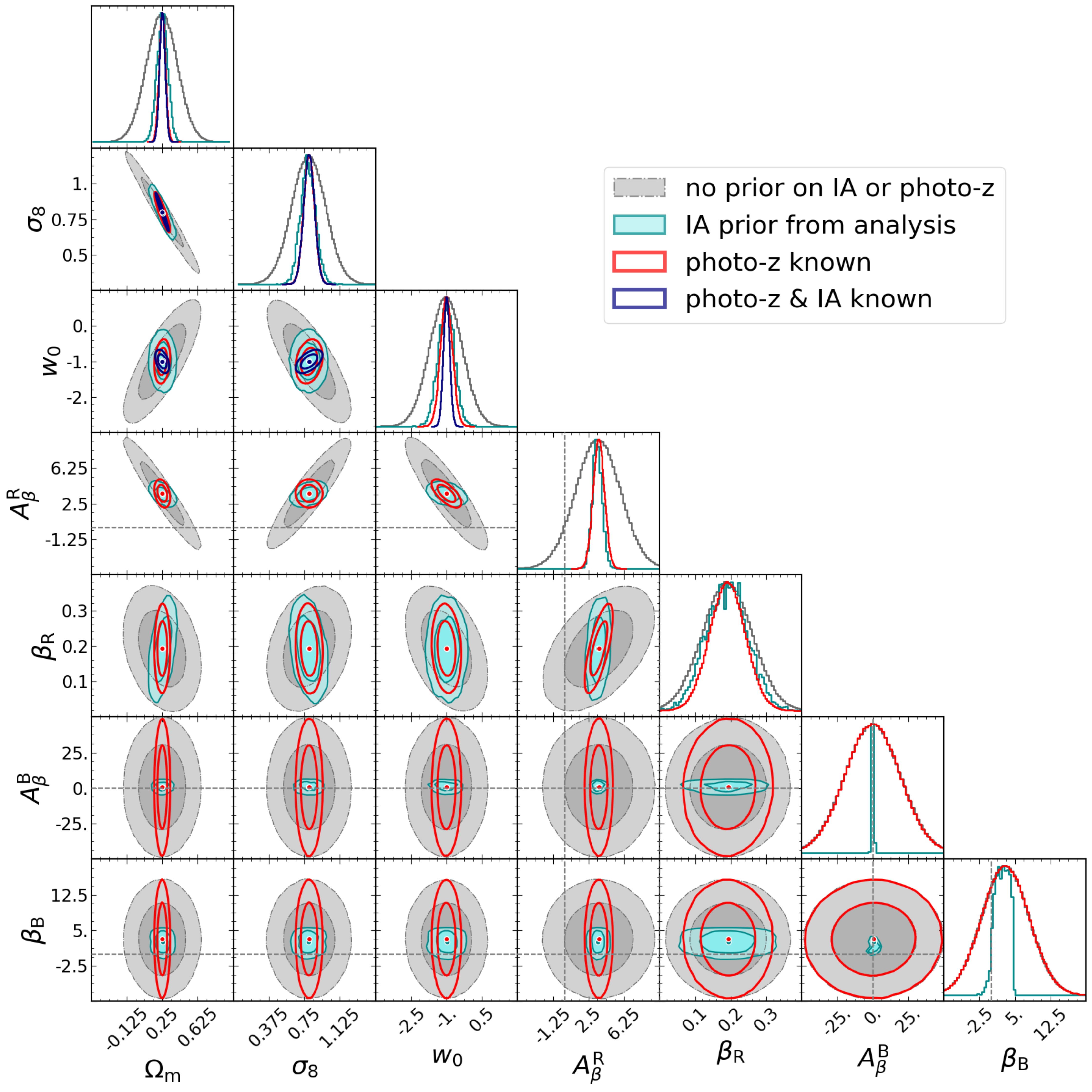}
    \caption{The same as Figure \ref{fig:forecast_nobeta}, but assuming luminosity-dependent non-linear alignments (NLA-$\beta$) in the data.}
    \label{fig:forecast_beta}
\end{figure*}

\begin{figure*}
	\includegraphics[width=\textwidth]{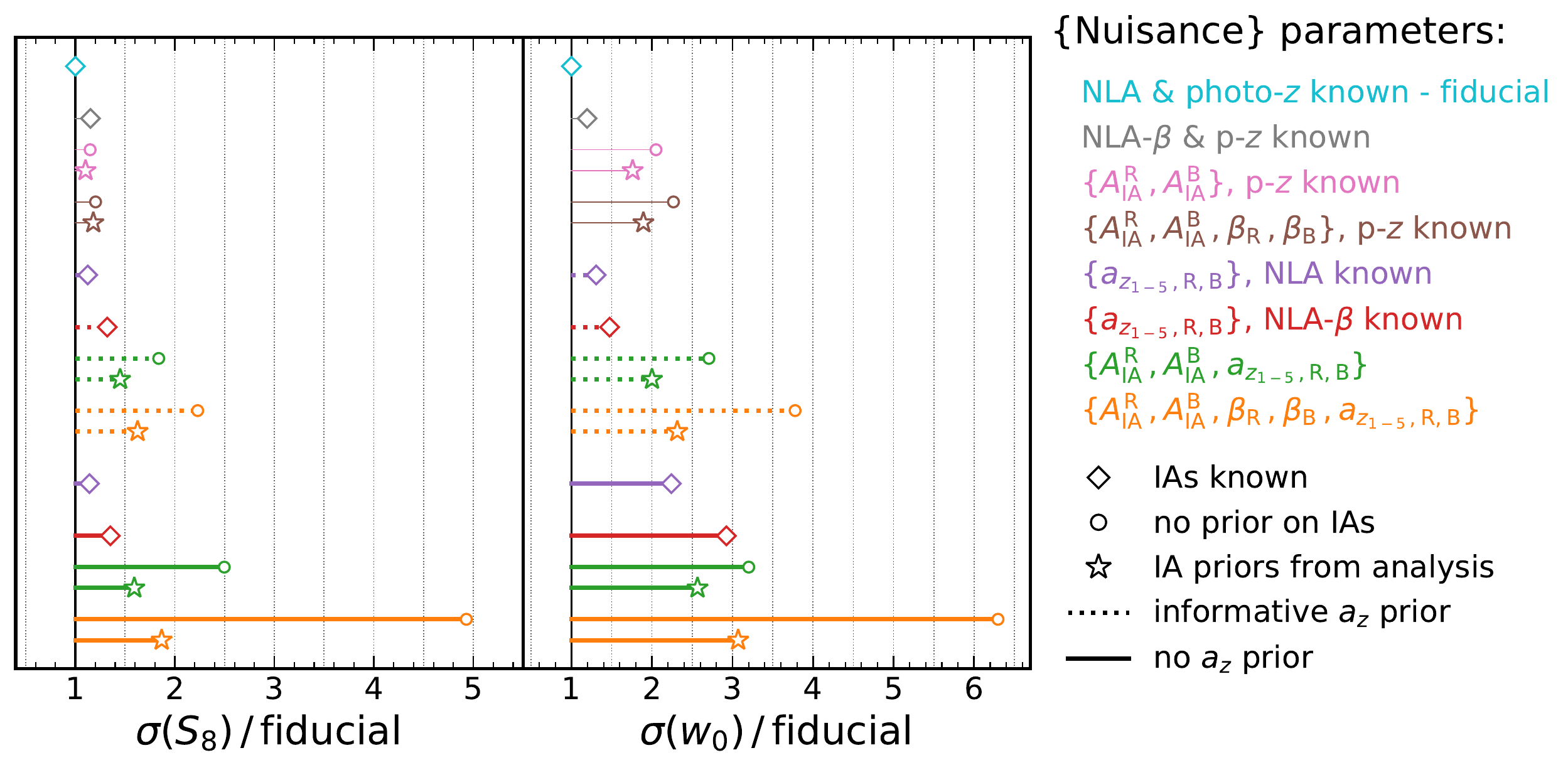}
    \caption{Comparison of marginalised $68\%$ confidence intervals on $S_{8} \equiv \sigma_{8}\sqrt{\Omega_{\rm{m}}/0.3}$ and $w_{0}$, forecasted for varied sets of nuisance parameters, with (stars) 
    and without (circles) the application of our intrinsic alignment parameter priors. Diamonds illustrate cases with perfect knowledge of the alignments in the simulated data. The intervals are plotted as ratios to the fiducial case (cyan), where NLA parameters and photo-$z$ distributions are perfectly known. Dotted lines denote cases with the `Informative' photo-$z$ bias priors (Table \ref{tab:forecast_priors}), and solid lines those without any prior. Nuisance (free) parameters are denoted for each case by curly brackets $\{\}$ in the legend.}
    \label{fig:forecast_constraints}
\end{figure*}

Here we forecast the impact of our informative IA priors upon a colour-split cosmic shear analysis over a completed KiDS survey. We assume that the alignments in the data are perfectly described by the N/LA models -- something we will investigate in future work. This assumption is questionable for the NLA, but given its widespread use in current surveys, and since we are not concerned with biasing of parameters here, but rather the pure impact of priors, we continue as such. The model survey is described by an area of $1,350\,\rm{deg}^{2}$ with a total galaxy number density of $9$ arcmin$^{-2}$ \citep{Hildebrandt2016}, and a total shape dispersion of $0.41$. We model the $n(z)$, over $z \in [\,0.1\,, 1.2\,]\,$, according to \citep{Smail1995}

\begin{equation}
	n_{\textrm{total}}(z) \propto z^{\alpha}\, \textrm{exp}\left\{ - \left( \frac{z}{z_{0}} \right) ^{\gamma} \right\} \quad ,
\end{equation}
where $\alpha=2$, $\gamma=1.5$ and $z_{0}=0.375$. We define $5$ tomographic bins in redshift, with edges (KiDS+VIKING-450; \citealt{Hildebrandt2018}): $[\,0.1\,,0.3\,,0.5\,,0.7\,,0.9\,,1.2\,]$, each scattered about the bin centre with $\sigma_{z}=0.05(1+z)$ and with no catastrophic outliers. Using KV450 galaxies, we estimate $\left\langle{}L/L_{\rm{piv}}\right\rangle$ for each colour/redshift bin, with the unchanged pivot $L_{\rm{piv}}\sim\num{4.6e10}L_{\sun}$. We also assume the KV450 red galaxy fraction per redshift bin for our toy survey; approx. $[\,0.13\,,0.23\,,0.27\,,0.26\,,0.26\,]$ \citep{Wright2018}. Splitting the model survey by colour more than doubles the available information when computing auto- and cross-correlations -- our data vector $\vec{d}$ consists of shear angular power spectra $C(\ell)$ with intrinsic contributions (Eqs. \ref{eq:shearspectrum}-\ref{eq:C_GI}), for all colour/redshift bin combinations, in $10$ logarithmic bins $\ell \in [\,50\,, 2000\,]$

\begin{equation}
  \begin{split}
	\vec{d} = \{\, & C_{i_{\textrm{\ssz{B}}}j_{\textrm{\ssz{B}}}}(\ell) \quad \forall \quad i \,, j \in [1\,,5] \, \cap \, i \leqslant j \quad , 	\\
	& C_{i_{\textrm{\ssz{B}}}j_{\textrm{\ssz{R}}}}(\ell) \quad \forall \quad i \,, j \in [1\,,5] \quad ,	\\
	&  C_{i_{\textrm{\ssz{R}}}j_{\textrm{\ssz{R}}}}(\ell) \quad \forall \quad i \,, j \in [1\,,5]\, \cap \, i \leqslant j \, \} \quad ,
  \end{split}
\end{equation}
for a total of $550$ data points. We compute a full analytical covariance matrix \citep[see][Section 5]{Hildebrandt2016}, with non-Gaussian and super-sample contributions, for computation of the Fisher information \citep*[see][and references therein]{Tegmark1997}. Our cosmological parameter vector is

\begin{equation}
	\vec{\lambda}^{\rm{Fisher}} = \{\,\Omega_{m}\,, \Omega_{b}\,, h\,, \sigma_{8}\,, n_{s}\,, w_{0}\,\} \quad ,
\end{equation}
and we fix $\Omega_{k}=0$. We append the parameter vector with nuisance parameters for the NLA/NLA-$\beta$, and for characterising the impact of additive photometric redshift biases -- modern shear surveys rely upon photo-$z$, and as such are prone to systematic bias in redshift distributions and resultant constraints. We parameterise the additive photo-$z$ bias per colour/redshift bin, such that $n_{\rm{x}}(z) \rightarrow n_{\rm{x}}(z\,-\,a_{z_{\rm{x}}})$ , where $n_{\rm{x}}(z)$ is the redshift distribution of bin $\rm{x} \in [\,1\,,5\,]_{\textrm{\ssz{R,B}}}$. Our nuisance parameters are then

\begin{equation}
	\{\Aia\,, \beta\,, a_{z_{1}}\,, a_{z_{2}}\,, a_{z_{3}}\,, a_{z_{4}}\,, a_{z_{5}}\,\}_{\textrm{\ssz{R,B}}} \quad ,
    \label{eq:forecast_params}
\end{equation}
giving a total of $18$ $(20)$ parameters with NLA (NLA-$\beta$) alignments in the data. We take the MICE cosmology from our IA analysis as the fiducial cosmology about which Fisher derivatives are computed, and apply Gaussian priors as listed in Table \ref{tab:forecast_priors}. Adding a Gaussian prior to the Fisher information is necessary in the case of $A^{\textrm{\ssz{B}}}_{\beta}$ and $\beta_{\textrm{\ssz{B}}}$, as small-amplitude signals result in a total degeneracy between these parameters. We limit their variability -- in the NLA-$\beta$ forecast, only -- in order to demonstrate a meaningful application of our derived IA priors. The results of our forecasts are shown in Figures \ref{fig:forecast_nobeta} \& \ref{fig:forecast_beta}, and condensed in Figure \ref{fig:forecast_constraints} to show the IA prior impacts on the $S_{8} \equiv \sigma_{8}\sqrt{\Omega_{\rm{m}}/0.3}$ parameter, and dark energy equation of state $w_{0}$. We note that the Fisher approximation -- the mean curvature of the likelihood function about the fiducial cosmology -- is inexact in the case of non-Gaussian posterior probability distributions, such that the banana-like $\Omega_{\rm{m}} - \sigma_{8}$ degeneracy observed in cosmic shear analyses (\citealt{Hildebrandt2016}, \citealt{DESCollaboration2017}) is not exactly captured. Thus our forecasts are demonstrative in purpose, and may differ from analogous full, simulated likelihood forecasts \citep*[e.g.][]{Krause2016}.

Figures \ref{fig:forecast_nobeta} \& \ref{fig:forecast_beta} depict forecasted constraints for final KiDS-like colour-split cosmic shear, with NLA and NLA-$\beta$ alignments in the modelled data, respectively. Clear improvements are seen in $\Omega_{m}\,,\sigma_{8}\,,w_{0}$ constraining power when applying our $68\%$ confidence intervals as informative priors on the IA parameters (grey vs. cyan contours). This demonstrates the degrading influence of free-to-roam IA nuisance parameters in cosmic shear analyses -- the application of our priors results in up to $\sim50\%$ reductions in the size of errorbars on $S_{8}$ and $w_{0}$ for the NLA forecast, and $20\%$ for the LA whose weaker contribution levies smaller gains when constrained. The gains in constraints upon crucial parameters are illustrated in Figure \ref{fig:forecast_constraints} (circles vs. stars) and fully detailed in Table \ref{tab:forecast_results}, for each of the photo-$z$ bias prior setups detailed in Table \ref{tab:forecast_priors}. The `Modest' case features a rough estimate for a monotonically increasing uncertainty in the real positions of tomographic bin-centres, and serves as a yard-stick between the case without any priors and the `Informative' case, where we adapt the constraints of \cite{VanUitert2018} for use as priors.

\begin{table*}
\begin{center}
	\caption{Forecasted improvements in constraining power for key cosmological parameters when employing IA model constraints as informative prior ranges on IA nuisance parameters. The variable parameters for each mode of an N/LA forecast are indicated by curly brackets \{\} in the left-most column. Improvements are given as the percentage reductions in the sizes of $1\sigma$ confidence intervals on respective parameters, for the no-prior, modest-prior and informative-prior photo-$z$ bias prior setups (see Table \ref{tab:forecast_priors}), hence we only display a single reduction for the cases without any photo-$z$ bias.}
	\label{tab:forecast_results}
	\def\arraystretch{1.6}
	\begin{tabular*}{\textwidth}{lcccc} 
    \hline
    \{Nuisance\} Parameters & $\Omega_{\rm{m}}$ & $\sigma_{8}$ & $S_{8} \equiv \sigma_{8}\sqrt{\Omega_{\rm{m}}/0.3}$ & $w_{0}$ \\
    \hline
    Linear alignments (LA) & & & & \\
    $\{A_{\rm{IA}}^{\rm{R}} \,, A_{\rm{IA}}^{\rm{B}}\}$, photo-$z$ known & 2\% & 1\% & 1\% & 1\% \\
$\{A_{\rm{IA}}^{\rm{R}} \,, A_{\rm{IA}}^{\rm{B}} \,, \beta_{\rm{R}} \,, \beta_{\rm{B}}\}$, photo-$z$ known & 6\% & 2\% & 4\% & 7\% \\
$\{A_{\rm{IA}}^{\rm{R}} \,, A_{\rm{IA}}^{\rm{B}} \,, a_{z_{1-5}\,,\rm{R,B}}\}$ & 3\%, 4\%, 3\% & 2\%, 1\%, 1\% & 2\%, 2\%, 2\% & 2\%, 1\%, 1\% \\
$\{A_{\rm{IA}}^{\rm{R}} \,, A_{\rm{IA}}^{\rm{B}} \,, \beta_{\rm{R}} \,, \beta_{\rm{B}} \,, a_{z_{1-5}\,,\rm{R,B}}\}$ & 27\%, 24\%, 20\% & 22\%, 17\%, 14\% & 25\%, 21\%, 17\% & 17\%, 22\%, 20\% \\
    \hline
    Non-linear alignments (NLA) & & & & \\
    $\{A_{\rm{IA}}^{\rm{R}} \,, A_{\rm{IA}}^{\rm{B}}\}$, photo-$z$ known & 7\% & 1\% & 4\% & 14\% \\
$\{A_{\rm{IA}}^{\rm{R}} \,, A_{\rm{IA}}^{\rm{B}} \,, \beta_{\rm{R}} \,, \beta_{\rm{B}}\}$, photo-$z$ known & 2\% & 0\% & 1\% & 15\% \\
$\{A_{\rm{IA}}^{\rm{R}} \,, A_{\rm{IA}}^{\rm{B}} \,, a_{z_{1-5}\,,\rm{R,B}}\}$ & 38\%, 28\%, 25\% & 27\%, 18\%, 14\% & 36\%, 25\%, 21\% & 20\%, 27\%, 26\% \\
$\{A_{\rm{IA}}^{\rm{R}} \,, A_{\rm{IA}}^{\rm{B}} \,, \beta_{\rm{R}} \,, \beta_{\rm{B}} \,, a_{z_{1-5}\,,\rm{R,B}}\}$ & 64\%, 38\%, 32\% & 56\%, 23\%, 17\% & 62\%, 36\%, 27\% & 51\%, 43\%, 39\% \\
    \hline
	\end{tabular*}
\end{center}
\end{table*}

Figures \ref{fig:forecast_nobeta}, \ref{fig:forecast_beta} \& \ref{fig:forecast_constraints} also plot some idealised cases -- assuming perfect knowledge of both intrinsic alignments and photometric redshift distributions we plot navy, unfilled contours in Figures \ref{fig:forecast_nobeta} \& \ref{fig:forecast_beta}, and cyan/grey diamonds in Figure \ref{fig:forecast_constraints}. For perfect photo-$z$ alone, we include red, unfilled contours (Figures \ref{fig:forecast_nobeta} \& \ref{fig:forecast_beta}) and mauve/red diamonds (Figure \ref{fig:forecast_constraints}). In the latter case, our analysis priors are not applied, as they have a negligible effect upon the constraining power of the model survey -- i.e. with perfect knowledge of source redshifts, such a survey could self-calibrate for intrinsic alignments beyond the precision of our direct analysis. The difference between the red/blue unfilled contours in Figures \ref{fig:forecast_nobeta} \& \ref{fig:forecast_beta} is then the potential gain in precision from even tighter IA model constraints, which is seen to be particularly large for the dark energy equation of state $w_{0}$. 

The advantages of informative priors on intrinsic alignment model parameters are clearly demonstrated here, especially when considering that the current modes of modelling are too simple -- our colour-split analysis is already more complex than most. Alignment models with additional freedoms must be used, in order to characterise the variable contributions of galaxies of different types and in different environments -- e.g. the mixed alignment perturbative model of \cite{Blazek2017}, recently applied to DES Y1 data \citep{Troxel2017} and accompanied by losses in constraining power. The mitigation of such losses demands dedicated IA studies, producing reliable priors for model parameters.

\section{CONCLUSIONS}
\label{sec:conclusions}

We have measured the galaxy position-intrinsic shear and position-position correlations in the GAMA and SDSS Main galaxy samples, selecting subsamples by colour and redshift. We undertook a detailed consideration of reliable subsample covariance estimation, implementing a 3-dimensional jackknife routine for the relatively small-area GAMA samples. We jointly fit to our intrinsic alignment and clustering measurements with several models; the non-linear and linear alignment models (N/LA), and luminosity-dependent analogues (N/LA-$\beta$).

Our NLA fits yield constraints (quoted to $1\sigma$) upon the intrinsic alignment amplitude $\Aia$ for 3 cases; unselected, early-type and late-type galaxies, each representing a step forward in precision for constraints of their type from dedicated, spectroscopic studies of intrinsic alignments. Our findings agree with the literature, wherein red galaxies exhibit significant, positive (radial) alignments, and blue galaxy alignments are thus far undetectable. We also fit the LA model to our data, finding comfortable consistency with each of our results for the NLA. As noted in the text, this is largely due to our restriction to linear scales $>6\mpc$ where the N/LA difference is minimal.

Our red galaxy alignment constraint $\Aia = 3.18^{+0.46}_{-0.45}$ appears to demonstrate that fainter, non-LRG galaxies are still privy to a radial alignment mechanism on large scales (up to $60\mpc$ in this analysis), although not as strongly as LRGs (e.g. \citealt{Joachimi2011}, \citealt{Singh2015}). We are able to improve constraints upon the blue galaxy alignment amplitude to $\Aia = 0.21^{+0.37}_{-0.36}$, consistent with the work of \cite{Mandelbaum2011}, and still consistent with a null signal. This result, from scales $>6\mpc$\footnote{The small scales neglected in fitting are similarly discarded in most $3\times2$pt analyses, due to uncertainties in the modelling of non-linearities and baryonic contributions.}, supports the quadratic alignment picture of weak spiral galaxy alignments on linear scales. Fitting jointly to the \wgp (and $w_{\rm{gg}}$) signals measured in GAMA and SDSS, without any colour or redshift selections, yields $\Aia=1.06^{+0.47}_{-0.46}$, signifying a net radial alignment of galaxies in the combined dataset.

In the context of contaminations to weak lensing, the result for blue galaxies may be the most pertinent -- whilst our flux-limited samples offer the most representative dataset we can muster, the difficulties of spectroscopy limit them to relatively bright galaxies at low redshifts. Thus our model constraints for the unselected case are likely to over-predict red galaxy contributions -- photometric cosmic shear datasets extend to greater depths and hence higher redshifts, where faint, blue galaxies dominate samples. While the results of our fitting to individual galaxy samples reveal weak/non-existent correlations between IA and galaxy luminosity/redshift, we also find significant, scale-dependent variability of IA when separating central/satellite contributions. 
In GAMA, red central and satellite galaxies align with their local galaxy distribution, i.e. that of the group halo, whilst red central shapes are solely responsible for the linear-scale correlation. Any blue central/satellite galaxy alignments remain undetectable. A full consideration is beyond the scope of this work, however our derived IA constraints remain the most representative for shear-like samples, and should be instructive for future studies.

We recommend the use of our colour-specific alignment constraints, and our interpolation between them (Eq. \ref{eq:redfrac}), in formulating a prior range on $\Aia$ for future cosmic shear signal fitting. An average of our constraints, weighted by the relative red/blue galaxy populations, is likely to provide a more realistic description of the alignments present in a dataset -- noting the GAMA satellite fraction of $\sim27\%$, one can consider the $\Aia$ interpolation to serve as a conservative upper-limit for similarly satellite-heavy samples. 

Our fits of the luminosity-dependent NLA resulted in null detections for the $\beta$ power-law, at $95\%$ confidence. The blue galaxy $\beta$ parameter is poorly constrained by the data, as the luminosity baseline of the samples is sparse and ineffectual, and the signals are close to zero. The red galaxy result is interesting, as it seems to contradict previous works which have found a roughly linear scaling of alignments with sample luminosity (\citealt{Joachimi2011}, \citealt{Singh2015}). 
The reason for this is that the red galaxy alignments show little/no evolution over a luminosity baseline of (rescaled to start at unity) $\sim \,[\,1\,,1.67\,,5\,]$. This observation might be partly explained by the density weighting and consequent bias-dependent signal enhancement described by \cite*{Blazek2015}, and certainly lends support to the notion that the current methods of modelling for IA are insufficient to grasp the complexity of contributions from galaxies in different environments.

We forecasted the cosmological parameter constraining capabilities of red/blue-split cosmic shear in a completed KiDS survey, assuming that alignments in the simulated data were described by the NLA or LA models. Applying our IA nuisance parameter constraints as informative priors, we find reductions of up to $\sim50\%$ (or $\sim20\%$ for the LA) in the size of confidence intervals for the $S_{8}$ parameter and $w_{0}$, dependent on the freedoms of photometric redshift bias parameters. Our forecasts demonstrate the potential utility of independent intrinsic alignment model constraints as informative priors in cosmic shear analyses, particularly as IA parameterisations become more complex and impactful.

In the era of LSST, Euclid and WFIRST, our current prescriptions for the intrinsic alignment contamination of cosmic shear would lag behind greatly increased statistical power -- one fears that the limit of cosmological inference could be determined by the uncertainty in models for IA (and other systematics), and open to strong biases as a result. Our work has attempted to characterise the alignment signatures of a purely flux-limited sample, finding complexity beyond the divergent behaviour of elliptical and spiral galaxies, extending to the non-trivial contributions of red centrals and satellites. These findings motivate us to explore IA models with galaxy red- and satellite-fraction considerations, and to constrain such models with representative spectroscopic data -- such work will aid in the maximisation of potential for the next generation of lensing surveys.

Looking forward, we hope to perform this analysis with a halo model for intrinsic alignments, adapted from the formalism of \cite{Schneider2010}, fitting to all scales, including a satellite-alignment prescription, and taking full advantage of the high completeness of these data (Fortuna et al., in prep.). In the meantime, our derived NLA model constraints will provide useful priors for current and future shear surveys, improving cosmological constraints and blocking the influence of unknown systematics on IA parameterisations. New, narrow-band photometric datasets are currently being amassed (PAUS; \citealt{Benitez2009}, J-PAS; \citealt{Benitez2014}), with the potential for the production of unprecedented IA model constraints, for use in future weak lensing analyses. Furthermore, the statistical power and associated precision of these datasets will enable the use of intrinsic-intrinsic shear (II) correlations in studying the type- and environment-dependence of galaxy alignment mechanisms. Powerful and additional statistics, cross-correlations between galaxy types, and increased depths in these analyses will shed new light on the intrinsic alignment contamination of cosmic shear, and on the physics of galaxy formation and evolution.



\section*{Acknowledgements}

We thank Rachel Mandelbaum and Sukhdeep Singh for sharing data products and insights. We also thank members of the KiDS consortium for many helpful discussions.

HJ acknowledges support from a UK Science \& Technology Facilities Council (STFC) Studentship. HH and MCF acknowledge support from Vici grant 639.043.512, financed by the Netherlands Organisation for Scientific Research (NWO). NEC acknowledges support from a Royal Astronomical Society Research Fellowship. CH acknowledges support from the European Research Council under grant number 647112. KK acknowledges support by the Alexander von Humboldt Foundation.

This work is based on data products from observations made with ESO Telescopes at the La Silla Paranal Observatory under programme IDs 177.A-3016, 177.A-3017 and 177.A-3018, and on data products produced by Target/OmegaCEN, INAF-OACN, INAF-OAPD and the KiDS production team, on behalf of the KiDS consortium.

We also work with data products from GAMA: a joint European-Australasian project based around a spectroscopic campaign using the Anglo-Australian Telescope. The GAMA input catalogue is based on data taken from the Sloan Digital Sky Survey and the UKIRT Infrared Deep Sky Survey. Complementary imaging of the GAMA regions is being obtained by a number of independent survey programmes including GALEX MIS, VST KiDS, VISTA VIKING, WISE, Herschel-ATLAS, GMRT and ASKAP providing UV to radio coverage. GAMA is funded by the STFC (UK), the ARC (Australia), the AAO, and the participating institutions. The GAMA website is http://www.gama-survey.org/.

Part of this work was made possible thanks to CosmoHub. CosmoHub has been developed by the Port d'Informaci\'{o} Cient\'{i}fica (PIC), maintained through a collaboration of the Institut de F\'{i}sica d'Altes Energies (IFAE) and the Centro de Investigaciones Energ\'{e}ticas, Medioambientales y Tecnol\'{o}gicas (CIEMAT). The work was partially funded by the ``Plan Estatal de Investigaci\'{o}n Cient\'{i}fica y T\'{e}cnica y de Innovaci\'{o}n'' program of the Spanish government.

{\small\textit{Author Contributions:} All authors contributed to the development and writing of this paper. The authorship list is given in two groups: the lead authors (HJ, CG, BJ, HH), followed by an alphabetical group including those who are key contributors to either the scientific analysis, or to the data products.}



\bibliographystyle{aa}
\bibliography{IApaperRefs.bib}{}



\appendix

\section{COVARIANCES}
\label{sec:app_covariances}

To quantify sample variance, one would ideally prefer to use many realisations of simulated data in estimating errors on a statistic. Unfortunately, there remain qualitative disagreements between the latest hydrodynamical simulations with respect to the form of late-type galaxy intrinsic alignments \citep{Tenneti2016}. More fundamental road-blocks are small volumes and a lack of multiple realisations of these simulations, making them unsuitable for covariance purposes, as yet. We prefer our IA measurement errors to come from the data. 

With an eye to include the largest possible transverse scales in our analysis, we implement a 3-dimensional delete-one jackknife. The jackknife covariance is estimated as

\begin{equation}
	\hat{\tens{C}}_{{\rm{jack}}} = \frac{N-1}{N} \, \sum_{\alpha=1}^{N} \, \, (\vec{w}^{\alpha} - \bar{\vec{w}})(\vec{w}^{\alpha} - \bar{\vec{w}})^{\rm{T}} \quad ,
\end{equation}
where $\vec{w}^{\alpha}$ is the signal of interest, as measured from jackknife sample $\alpha$, and $\bar{\vec{w}}$ is the average over $N$ samples. $\rm{T}$ denotes the conjugate transpose of the mean-subtracted signal vector. $N$ jackknife samples are defined by dividing the survey into $N$ subvolumes and excluding one at a time, measuring $\vec{w}^{\alpha}$ in the rest of the survey. The performance of jackknife covariance estimation relies on a balance between (i) the number of jackknife subvolumes $N$, and (ii) the angular scale of their corresponding `patches' in the $\rm{RA}$-$\rm{DEC}$ plane, where one always compromises the other. $N$ should be $\gg$ the size of the data vector, or else the covariance becomes noisy and eventually singular. And yet, the scale of the subvolumes must be greater than the largest scales of interest, or the variance over those scales will not be captured and errors will be underestimated.

\begin{figure}
	\includegraphics[width=\columnwidth]{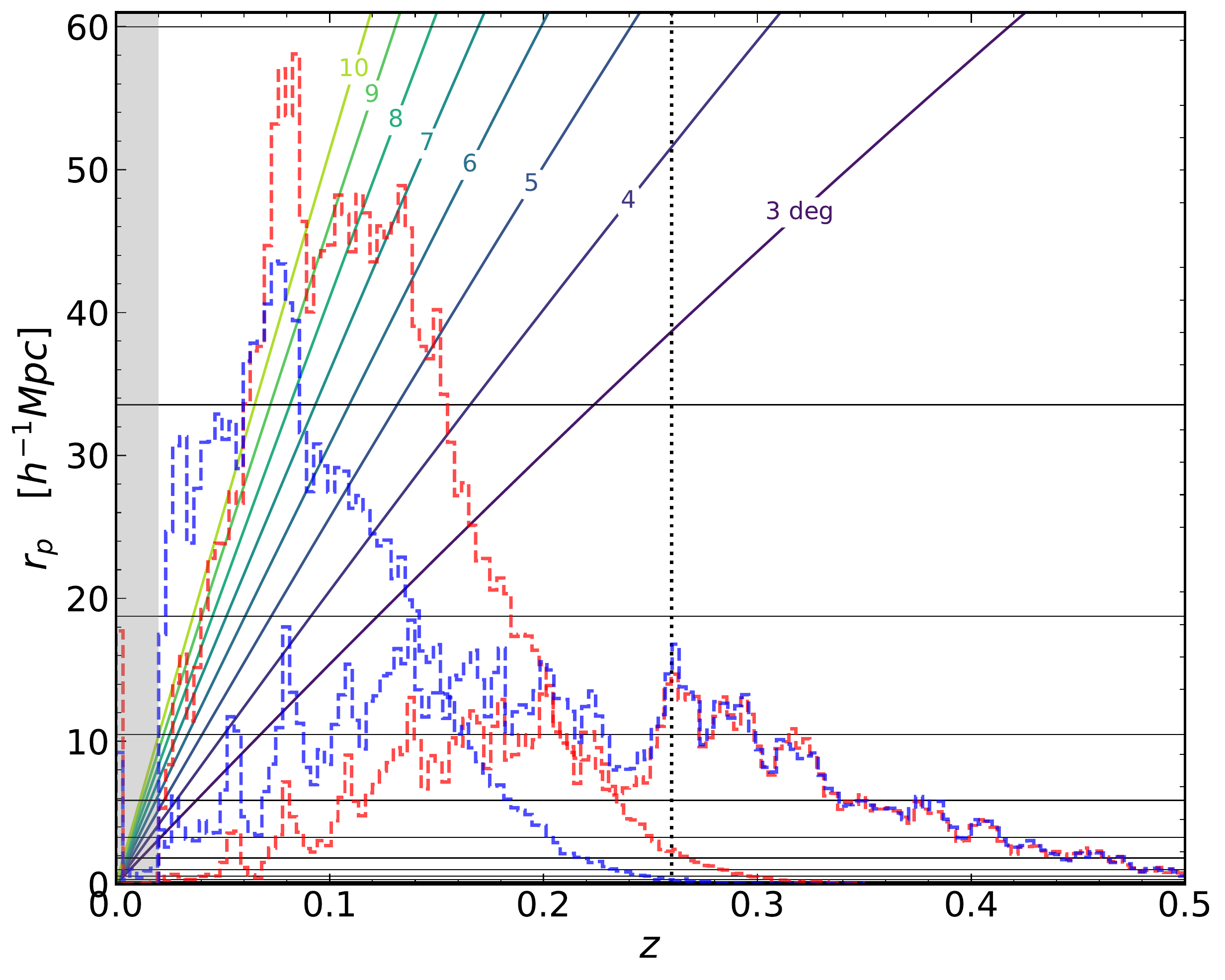}
    \caption{As a function of redshift, the comoving transverse vs. on-sky angular scale relation, for a range of scales in degrees. The $r_{p}$ bin edges we employ are plotted as horizontal black lines, highlighting the limitations of too-small patches to sample larger $r_{p}$ pairs, esp. at lower redshifts. The vertical dotted line indicates the $z=0.26$ redshift division for our GAMA samples, and redshift distributions of red and blue galaxies in SDSS ($z\sim [\,0.02\,, 0.3\,]$) and GAMA ($z\sim [\,0.02\,, 0.5\,]$) are overlaid as coloured, dashed histograms. Grey shading indicates $z < 0.02$, which we exclude from our analysis.}
    \label{fig:patchsizes}
\end{figure}

Figure \ref{fig:patchsizes} illustrates the mapping from an angular scale on the sky to a comoving transverse scale at a given redshift, with our log-spaced $r_{p}$-bin edges shown as horizontal black lines. This plot is interpreted as follows: coloured lines give the maximum comoving separations captured by an angular scale, thus anything below each line is correctly sampled by a sky-patch of that size, at that redshift. Normalised redshift distributions of GAMA and SDSS Main are overlaid, along with the redshift boundary defining our high- and low-$z$ GAMA samples (vertical red line). At high redshift, the GAMA jackknife demands patches of scale $\gtrsim4.5$ degrees for all $r_{p}$ scales to be captured, whilst the lower redshifts, which include all of the SDSS sample, are significantly hamstrung by the jackknife requirement.  Contiguous regions of equatorial GAMA are $12\,\rm{deg}\,\times\,5\,\rm{deg}$ in size, rendering ideal patches too few in number. SDSS Main covers a much larger area, but requires even larger patches at lower redshift. Thus we define a series of redshift slices, each with comparable numbers of galaxies, and subdivide jackknife patches into `cubes'. We take care to ensure that the resulting cubes are of more than sufficient depth to be considered statistically independent, and to accomodate the largest line-of-sight separations under consideration -- $|\Pi_{\rm{max}}| = 60\mpc$, so we ensure that all cubes are deeper than $150\mpc$. This requirement, along with the need for many jackknife regions of roughly equal galaxy numbers, is what informs our GAMA samples' shared redshift boundary at $z=0.26$. The large-$\Pi$ systematics test extends to $|\Pi_{\rm{max}}| = 90\mpc$, and we opt for a standard $2\rm{D}$ jackknife in this case, reducing the binning of the measured signal in order to stabilise the covariance matrix.

We note that by slicing subvolumes in redshift, we are assuming that we can approximate the variance over a redshift bin by the combined variance of its sub-bins, and thus that any redshift evolution is subdominant to the variance over different pointings. Since previous studies (\citealt{Joachimi2011}, \citealt{Mandelbaum2011}, \citealt{Tonegawa2017}) support weakly- or non-evolving alignments -- albeit for differently selected samples -- and our redshift baseline is short, we believe this assumption is reasonable.

\begin{figure*}
	\includegraphics[width=\textwidth]{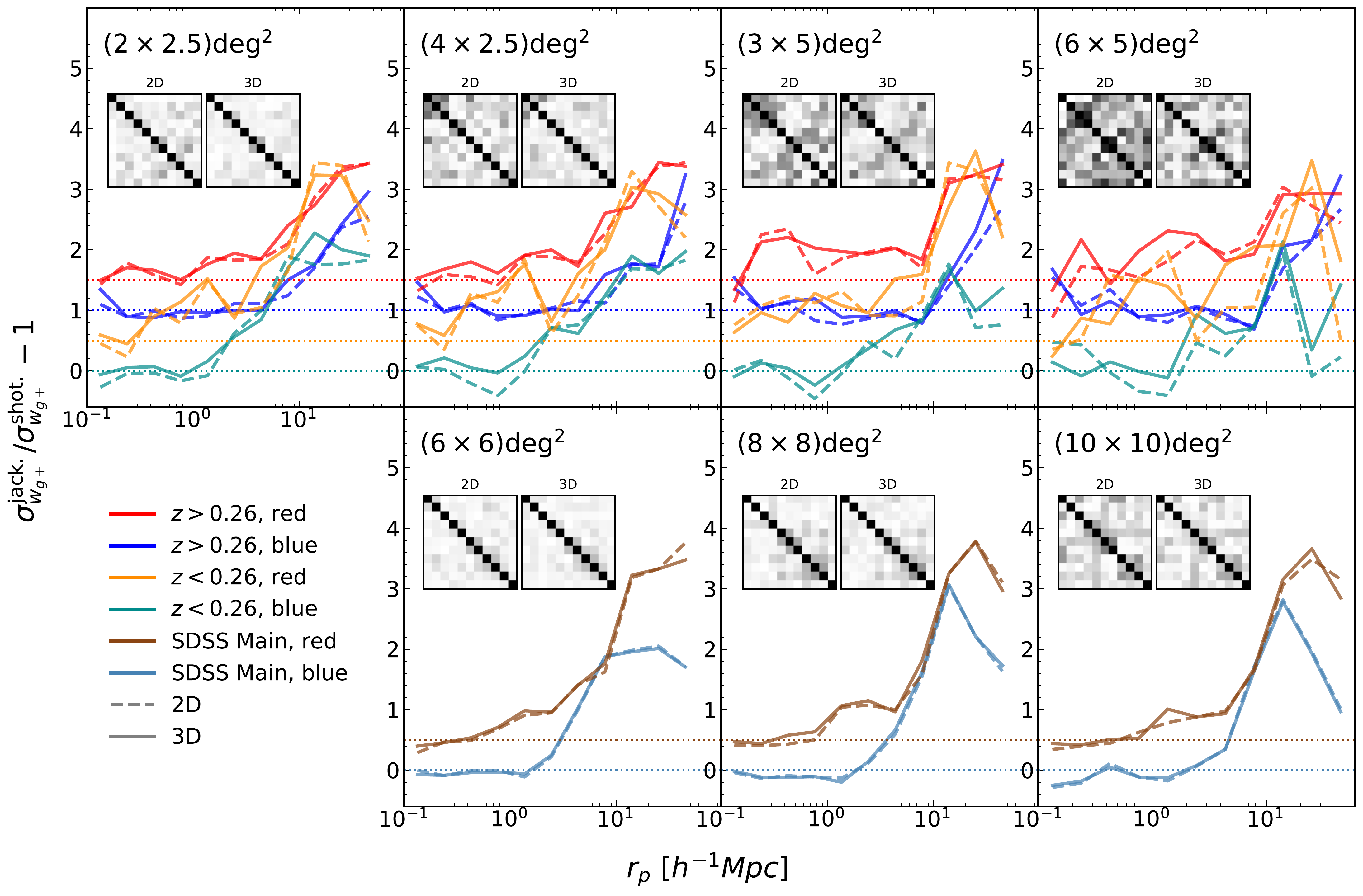}
    \caption{Performance of the $3\rm{D}$ (solid lines) and $2\rm{D}$ (dashed lines) jackknife in various configurations, plotted as ratios to analytical (shot-noise) errors estimated for the \wgp statistic. GAMA configurations are plotted on the top row, with SDSS on the bottom. Colour/redshift sample error ratios are vertically offset by increments of $0.5$ (equivalent to a $50\%$ difference in the error on \wgp) for clarity, and $\sigma^{\rm{jack.}}/\sigma^{\rm{shot.}}=1$ is indicated for each sample by a dotted horizontal line.
    Also shown are the Z2R and SR samples' $2\rm{D}$ (\emph{left-in-panel}) and $3\rm{D}$ (\emph{right-in-panel}) jackknife estimates of absolute correlation matrices $|R_{ij}|=|C_{ij}|\,\,/\sqrt{C_{ii}C_{jj}}$, for covariance $C_{ij}$, with $i\,,j \in [\,1\,,11\,]$ for $11$ bins in $r_{p}$.
    The angular dimensions $(\rm{RA}\times\rm{DEC})$ of jackknife patches, in degrees, are indicated.   
    Clearly visible trends are increasingly noisy covariances from larger/fewer patches, and the tendency of the $3\rm{D}$ jackknife to smooth this noise.
    The jackknife configurations we employ in our likelihood analysis are $(3\times5)\,\rm{deg}^{2}$ and $(10\times10)\,\rm{deg}^{2}$ for GAMA and SDSS, respectively.}
    \label{fig:3djackknife}
\end{figure*}

Figure \ref{fig:3djackknife} compares the performance of various $2\rm{D}$ and $3\rm{D}$ jackknife configurations. For our GAMA intrinsic alignment measurements we choose to work with $(3\times{}5)\,\rm{deg}^{2}$ patches, sliced into cubes -- the performance of this configuration is indicated by solid lines in the top-middle-right panel of Figure \ref{fig:3djackknife}. Subvolumes of this size combat the noise evident for $(6\times{}5)\,\rm{deg}^{2}$ patches (top-right panel), and remain large enough to sample all but the largest transverse scales at low redshifts -- we opt to drop only the largest-$r_{p}$ data point for low-redshift measurements. 
We describe our method for estimating clustering covariances in GAMA in Appendix \ref{sec:app_mice_errors}. For SDSS, we estimate both IA and clustering covariances with a $(10\times{}10)\,\rm{deg}^{2}$ jackknife cube configuration (Figure \ref{fig:3djackknife}, bottom-right, solid lines) -- the largest scales allowing for acceptable numbers of patches in the irregular SDSS footprint. We choose to retain all data points for SDSS measurements. 

Our chosen configurations yield $36\,,24$ and $74$ jackknife cubes per low-redshift GAMA, high-redshift GAMA and SDSS sample, respectively. Large-$\Pi$ test jackknife errors are derived from $12$ and $37$ patches ($2\rm{D}$) for GAMA and SDSS, respectively.

\subsection{Masking}
\label{sec:app_jackknife_masking}

While patches are chosen to be roughly equal in area, this is not always achieved due to masking and irregular survey edges. A patch covering less area translates into a less variant jackknife sample upon deletion. Thus when estimating the covariance from jackknife measurements, the noise at large scales is spuriously lowered, and inter-bin correlations are biased.

\benj{
We quantify this effect using data from the MICE Simulation. The Marenostrum Institut de Ci\'encies de l'Espai (MICE) Grand Challenge galaxy catalogue (\citealt{Carretero2015}, \citealt{Hoffmann2015}) was assembled from a $\num{7e10}$ dark matter particle, $\sim(3\,h^{-1}\rm{Gpc})^{3}$ comoving volume simulation \citep{Fosalba2015}, with halo occupation and abundance matching techniques \citep{Crocce2015}. The resulting catalogue spans a $5000\,\rm{deg}^{2}$ octant, complete down to an absolute $r$-band magnitude of $M_{r}<-18.9$.
}

\benj{
For a GAMA-sized patch of MICE, we generate a random ellipticity distribution and mask-out chunks of area in a similar fashion to the real masking in our KiDS images, estimating the jackknife alignment covariance before and after masking. We find that, whilst off-diagonal covariance elements can be severely mis-estimated, on-diagonal elements are recovered at $\sim23\%$ or better. We can lower this margin of error -- with a particular impact on the larger scales we use in fitting -- to $\sim18\%$ or better by applying weights to jackknife samples, equal to the relative areas of their respective deleted subvolumes (we apply an approximate re-normalisation incorporating the weights). 
}

\benj{
To test whether the more serious mis-estimation of off-diagonal covariance elements biases our results significantly, we set them all to zero and repeat our likelihood analysis. In comparison with our results quoted in Table \ref{tab:constraints}, we find consistency at $68\%$ confidence in all cases, with our fitted LA parameter values shifting as follows; $\{\Aiab:-1.2\sigma \,\,,\quad \Aiar:+0.33\sigma \,\,,\quad \Aia^{\textrm{full}}:+0.30\sigma\}$. Whilst red- and all-galaxy amplitude shifts are small at just $\sim0.3\sigma$, the larger, negative shift of $\sim1.2\sigma$ in the blue-galaxy amplitude fit acts to strengthen consistency with zero. For the LA-$\beta$, the red- and all-galaxy $\beta$ parameters shift to centre on zero, with small shifts taking the amplitudes toward the 1-parameter LA centres. The blue-galaxy $\beta$ parameter is relatively unchanged, with the amplitude centre shifting close to zero. Since none of these shifts contradict our original findings, the omission of inter-bin correlations can be said not to affect the conclusions of this work. We know that the worst biases of the off-diagonal covariance in our MICE test were equivalent to shifts in correlation coefficients of $\lesssim0.2$, thus we further conclude that biases of the signal covariance due to survey masking are subdominant to statistical errors for these data.
}

\subsection{GAMA clustering covariance}
\label{sec:app_mice_errors}

We make further use of MICE in estimating clustering covariances for GAMA -- for consistency, we impose the MICE faint-limit ($M_{r} \leqslant -18.9$) on our GAMA density samples, for losses of $\{\,\rm{Z1B}:27.3\% \,,\, \rm{Z1R}:5.6\% \,,\, \rm{Z2B}:0.1\% \,,\, \rm{Z2R}:0.1\%\,\}$ -- see Table \ref{tab:sample_details} for sample details. 

\begin{figure*}
	\includegraphics[width=\textwidth]{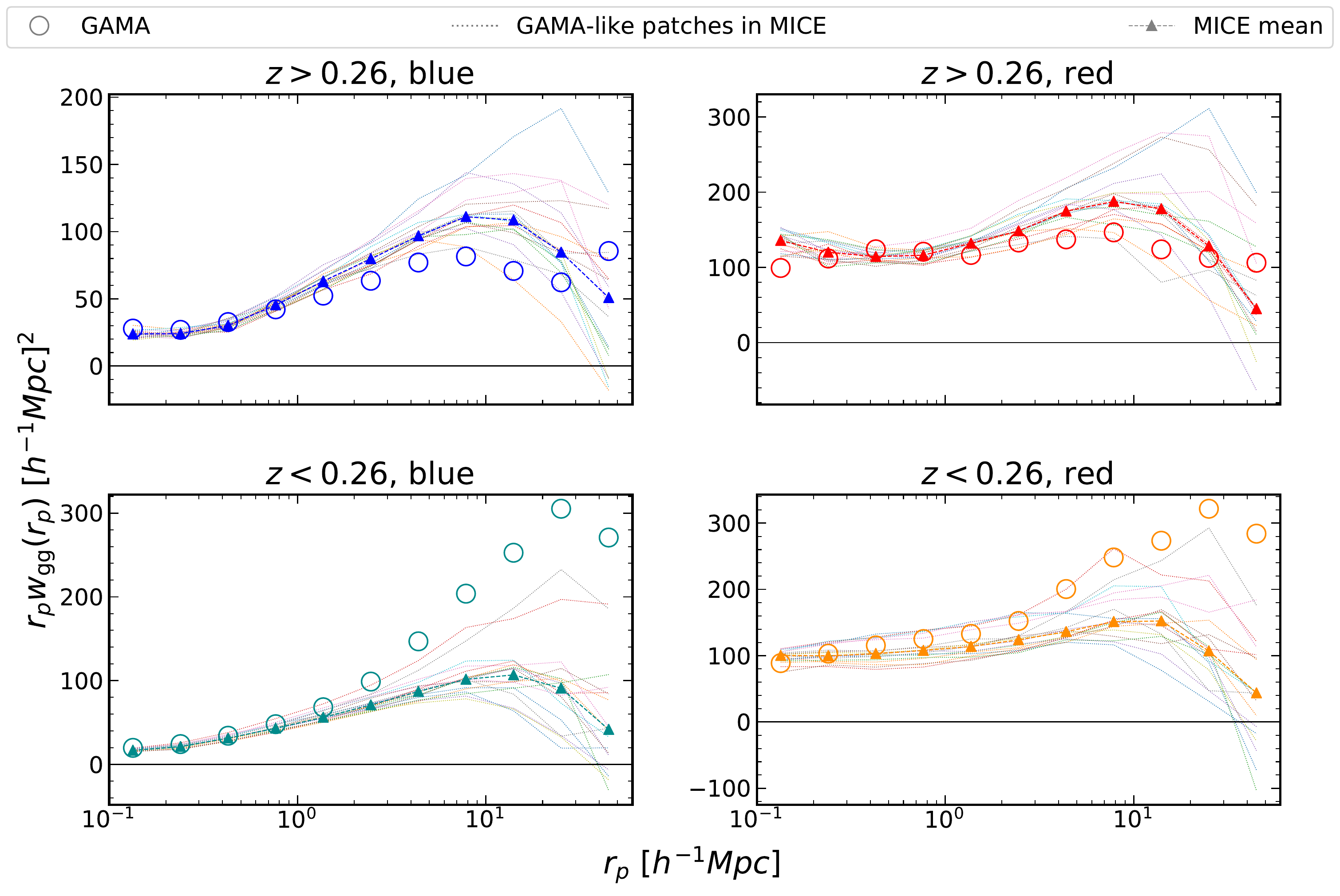}
    \caption{Clustering measurements from our defined GAMA galaxy samples (open circles) overlaid with corresponding measurements from individual, $\sim180\,\rm{deg}^{2}$ MICE subvolumes (dotted lines). Filled triangles show the means of the MICE clustering signals. We see significant differences between MICE and GAMA, particularly at low redshift and large scales, and so choose not to estimate covariances directly from MICE -- see Appendix \ref{sec:app_mice_errors} for details.}
    \label{fig:MICE_GAMA}
\end{figure*}

We apply the GAMA flux-limit $r<19.8$ to MICE and make a flat cut in absolute rest-frame $g-r$ to isolate the red sequence. Dividing the MICE area (with declination $\leqslant40\,\rm{deg}$) into $18$ rectangular patches, each $\sim{}180\,\rm{deg}^{2}$, we measure the clustering signals in each patch and find the spread to be slightly disagreeable with the clustering of analogous samples in GAMA (Figure \ref{fig:MICE_GAMA}). Thus we choose to validate the {\sc{swot}} clustering jackknife routine ({\sc{swot}}-jk), and its sensitivity to the jackknife subvolume numbers-vs.-size trade-off, using MICE.

\begin{figure*}
	\includegraphics[width=\textwidth]{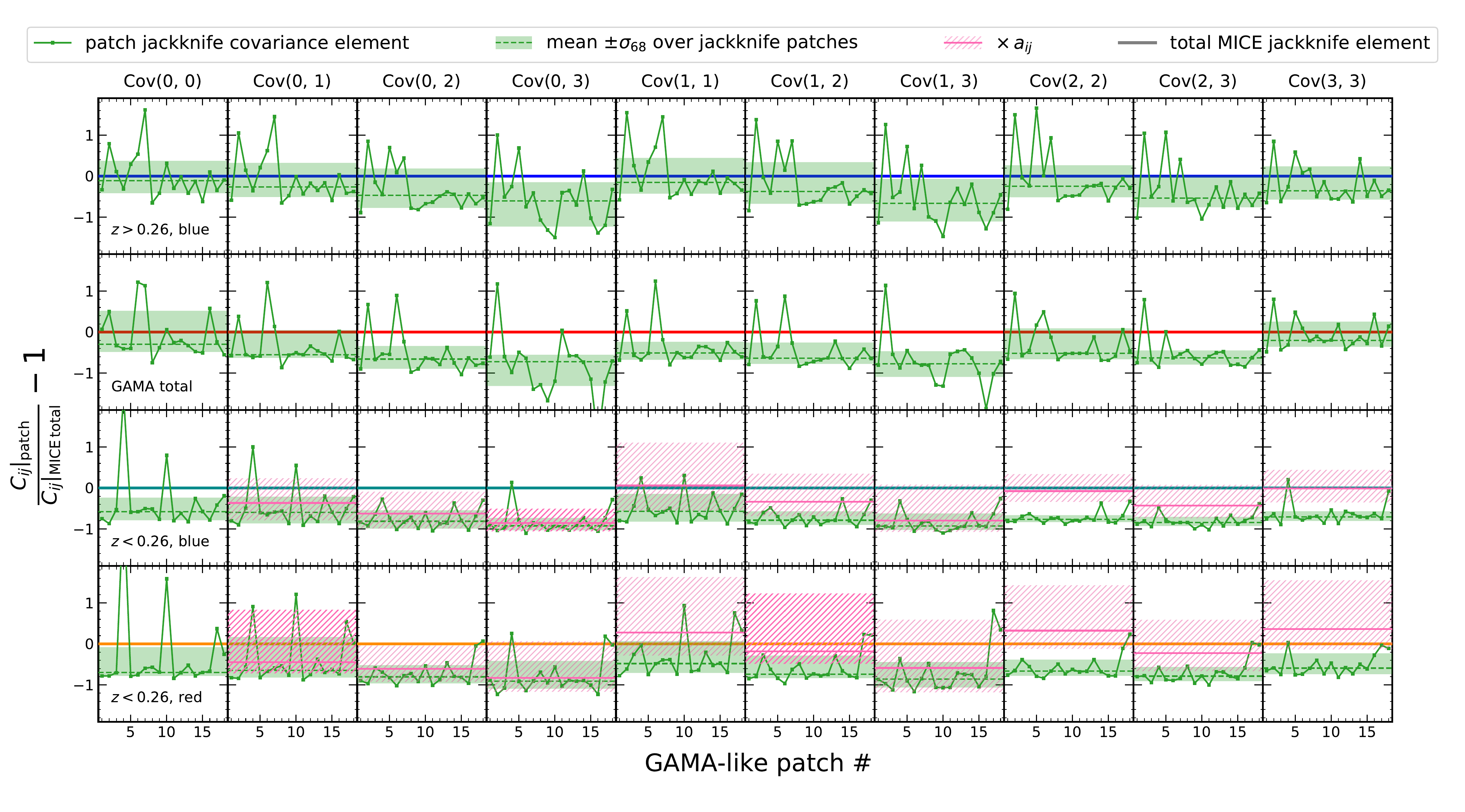}
    \caption{A comparison of clustering covariance elements (\emph{columns}), estimated for each redshift/colour galaxy sample (\emph{rows}) with the {\sc{swot}} internal jackknife per MICE patch ($X$, green) and the total jackknife over all patches ($Y$, solid coloured lines), plotted as $X/Y- 1$. The $0.16\,,0.5\,,0.84$ percentiles over the patch estimates are indicated by dashed lines and shading. Pink solid lines and hatching indicate covariance elements scaled by the fitted variables $a_{ij}$ described in Appendix \ref{sec:app_mice_errors}.}
    \label{fig:mice_covs}
\end{figure*}

We first obtain {\sc{swot}}-jk estimates of the MICE sample clustering signals and covariances, per patch. We then estimate total MICE sample clustering covariances by constructing $2\rm{D}$ jackknifes with all $18$ patches. The variance\footnote{We are now discussing the variance over independent estimates of the clustering covariance.} over {\sc{swot}}-jk estimates then approximates the sample variance of a $180\,\rm{deg}^2$ -- i.e. GAMA-like -- survey. If this is greater than any systematic offset between the mean covariance over the patches and the (area-scaled) total jackknife covariance, then the bias of the {\sc{swot}}-jk is subdominant to the statistical error of a GAMA-like survey. This is indeed the case for smaller transverse pair separations, which are well-sampled even by small angular scales. Samples at high redshift also do well in this regard, as smaller angular scales trace large spatial volumes. As Figure \ref{fig:mice_covs} illustrates, however, the {\sc{swot}}-jk significantly underestimates large-scale covariance elements for low-redshift samples (bottom panels; green lines \& shading vs. solid coloured lines). As discussed in Appendix \ref{sec:app_covariances} above, this is due to poor sampling of these pairs, and thus diminished variation across jackknife samples.

Since the jackknife performance differential is dominated by sample redshift, rather than colour, we attempt to quantify the lost variance by fitting 1 scaling variable to each of the $3$ largest transverse separation bins $i$ under consideration. Boosting each covariance element with the product $a_{ij}$ of the $2$ relevant scaling variables, we are able to bring the mean-over-patches (pink lines \& hatching in Fig. \ref{fig:mice_covs}) into closer agreement with the total MICE jackknife. We take these scaling factors to be approximately representative of the large-scale performance drop-off inherent to the {\sc{swot}}-jk at low redshift, and apply them to our low-$z$ GAMA clustering covariances. 

\section{INDIVIDUAL SAMPLE FITS}
Table \ref{tab:indiv_constraints} details the individual fits of the 1-parameter NLA model to galaxy samples, as described in Section \ref{sec:indivAia}, along with the relevant sample properties displayed in Figure \ref{fig:indiv_aia_plot}, and constraints upon the galaxy biases of corresponding density samples.

\begin{center}
\begin{table}
	\small
	\caption{1D marginalised constraints (to $1\sigma$) upon the 1-parameter NLA amplitude $\Aia$ for each sample under consideration -- each of the points shown in Figure \ref{fig:indiv_aia_plot} corresponds to a row here. `G' and `S' denote GAMA and SDSS samples, respectively. The independent galaxy samples whose signals constrain the `All blue' or `All red' amplitudes are denoted with $\dagger$. Samples denoted `high-$M_{*}$' are selected from GAMA to have stellar masses $>10^{11}M_{\sun}$. Also shown are red galaxy fractions $f_{{\rm{red}}}$, mean redshifts and mean luminosities (relative to the pivot $L_{{\rm{piv}}}=\num{4.6e10}L_{\sun}$) per shapes sample, and marginalised galaxy bias fits to corresponding density samples. Rows without these numbers correspond to amplitudes which were fit to multiple \wgp signals from samples with different properties.}
	\label{tab:indiv_constraints}
	\def\arraystretch{1.5}
	\begin{tabular}{lccccc} 
		\hline
		 Sample & $f_{{\rm{red}}}$ & $\langle{}z\rangle$ & $\langle{}L/L_{\rm{piv}}\rangle$ & $b_{\rm{g}}$ & $\Aia$ \\
		\hline
		\hline
G+S: full  &  0.54  &  --  &  --  &  --  &  $1.06^{+0.47}_{-0.46}$  \\
G: full  &  0.43  &  0.23  &  0.51  &  $1.57^{+0.08}_{-0.08}$  &  $0.26^{+0.63}_{-0.62}$  \\
S: full  &  0.60  &  0.11  &  0.22  &  $0.91^{+0.11}_{-0.12}$  &  $2.01^{+0.79}_{-0.71}$  \\
All blue  &  0  &  --  &  --  &  --  &  $0.21^{+0.37}_{-0.36}$  \\
G: high-$M_{*}$, blue  &  0  &  0.36  &  2.69  &  $1.62^{+0.11}_{-0.11}$  &  $2.72^{+2.54}_{-2.60}$  \\
$^{\dagger}$G: $z>0.26$, blue  &  0  &  0.33  &  1.06  &  $1.10^{+0.07}_{-0.07}$  &  $0.78^{+0.55}_{-0.54}$  \\
$^{\dagger}$G: $z<0.26$, blue  &  0  &  0.15  &  0.21  &  $1.55^{+0.09}_{-0.08}$  &  $-1.26^{+0.75}_{-0.69}$  \\
$^{\dagger}$S: blue  &  0  &  0.09  &  0.14  &  $0.88^{+0.12}_{-0.14}$  &  $1.03^{+0.90}_{-0.85}$  \\
All red  &  1  &  --  &  --  &  --  &  $3.18^{+0.46}_{-0.45}$  \\
G: high-$M_{*}$, red  &  1  &  0.31  &  2.03  &  $1.93^{+0.09}_{-0.10}$  &  $6.27^{+0.98}_{-0.96}$  \\
$^{\dagger}$G: $z>0.26$, red  &  1  &  0.33  &  1.47  &  $1.52^{+0.11}_{-0.11}$  &  $3.55^{+0.90}_{-0.82}$  \\
$^{\dagger}$G: $z<0.26$, red  &  1  &  0.17  &  0.50  &  $1.84^{+0.12}_{-0.12}$  &  $3.63^{+0.79}_{-0.79}$  \\
$^{\dagger}$S: red  &  1  &  0.12  &  0.29  &  $1.19^{+0.11}_{-0.11}$  &  $2.50^{+0.77}_{-0.73}$  \\
		\hline
	\end{tabular}
\end{table}
\end{center}

\section{LINEAR ALIGNMENT MODEL FITS}

Here we present fits of the linear alignment model to our IA data. Table \ref{tab:la_constraints} shows the results of our LA fitting, which are discussed in Section \ref{sec:la_results}.

\begin{table*}
 \small
 \begin{center}
 	\label{tab:la_constraints}
 	\caption{The same as Table \ref{tab:constraints}, here for the parameters of the linear alignment (LA) model (Section \ref{sec:nla}) and its luminosity-dependent analogue (LA-$\beta$).}
 	\def\arraystretch{1.5}
 	\begin{tabular*}{\textwidth}{lccclcccccc} 
 		\hline
 		 Sample & $\langle{}z\rangle$ & $\langle{}L/L_{*}\rangle$ & $b_{\rm{g}}$ & $\quad \quad \Aia$ & $\chi^{2}_{\nu}$ & $p(>\chi^{2})$ & $A_{\beta}$ & $\beta$ & $\chi^{2}_{\nu}$ & $p(>\chi^{2})$ \\
 		\hline
 		\hline
 GAMA full & 0.23 (0.24) & 0.51 (0.70) & $1.56_{-0.08}^{+0.09}$ & \rdelim\}{2}{1mm} \,\, \multirow{2}{*}{$1.23_{-0.52}^{+0.57}$} & \multirow{2}{*}{1.48} & \multirow{2}{*}{0.14} & \multirow{2}{*}{$1.37_{-2.45}^{+4.09}$} & \multirow{2}{*}{$2.25_{-2.53}^{+1.95}$} & \multirow{2}{*}{1.54} & \multirow{2}{*}{0.13} \\
SDSS Main full & 0.11 (0.11) & 0.22 (0.22) & $0.94_{-0.11}^{+0.10}$ &   &   &   &   &   &   &   \\
G: $z>0.26$, blue & 0.33 (0.33) & 1.06 (1.09) & $1.10_{-0.07}^{+0.07}$ & \rdelim\}{3}{1mm} &   &   &   &   &   &   \\
G: $z<0.26$, blue & 0.15 (0.17) & 0.21 (0.36) & $1.55_{-0.08}^{+0.09}$ & $\quad 0.35_{-0.43}^{+0.45}$ & 1.35 & 0.15 & $0.80_{-0.60}^{+0.55}$ & $2.57_{-1.69}^{+1.59}$ & 1.34 & 0.17 \\
S: blue & 0.09 (0.09) & 0.14 (0.14) & $0.88_{-0.15}^{+0.13}$ &   &   &   &   &   &   &   \\
G: $z>0.26$, red & 0.33 (0.33) & 1.47 (1.48) & $1.52_{-0.12}^{+0.10}$ & \rdelim\}{3}{1mm} &   &   &   &   &   &   \\
G: $z<0.26$, red & 0.17 (0.18) & 0.50 (0.56) & $1.85_{-0.13}^{+0.13}$ & $\quad 3.70_{-0.53}^{+0.52}$ & 1.34 & 0.16 & $3.93_{-0.63}^{+0.66}$ & $0.17_{-0.20}^{+0.20}$ & 1.40 & 0.14 \\
S: red & 0.12 (0.12) & 0.29 (0.29) & $1.19_{-0.12}^{+0.11}$ &   &   &   &   &   &   &   \\
 		\hline
 	\end{tabular*}
 \end{center}
\end{table*}

\section{PHOTO-Z BIAS PARAMETER CONTOURS}
\label{sec:app_photozbias_contours}

\begin{figure*}
	\includegraphics[width=\textwidth]{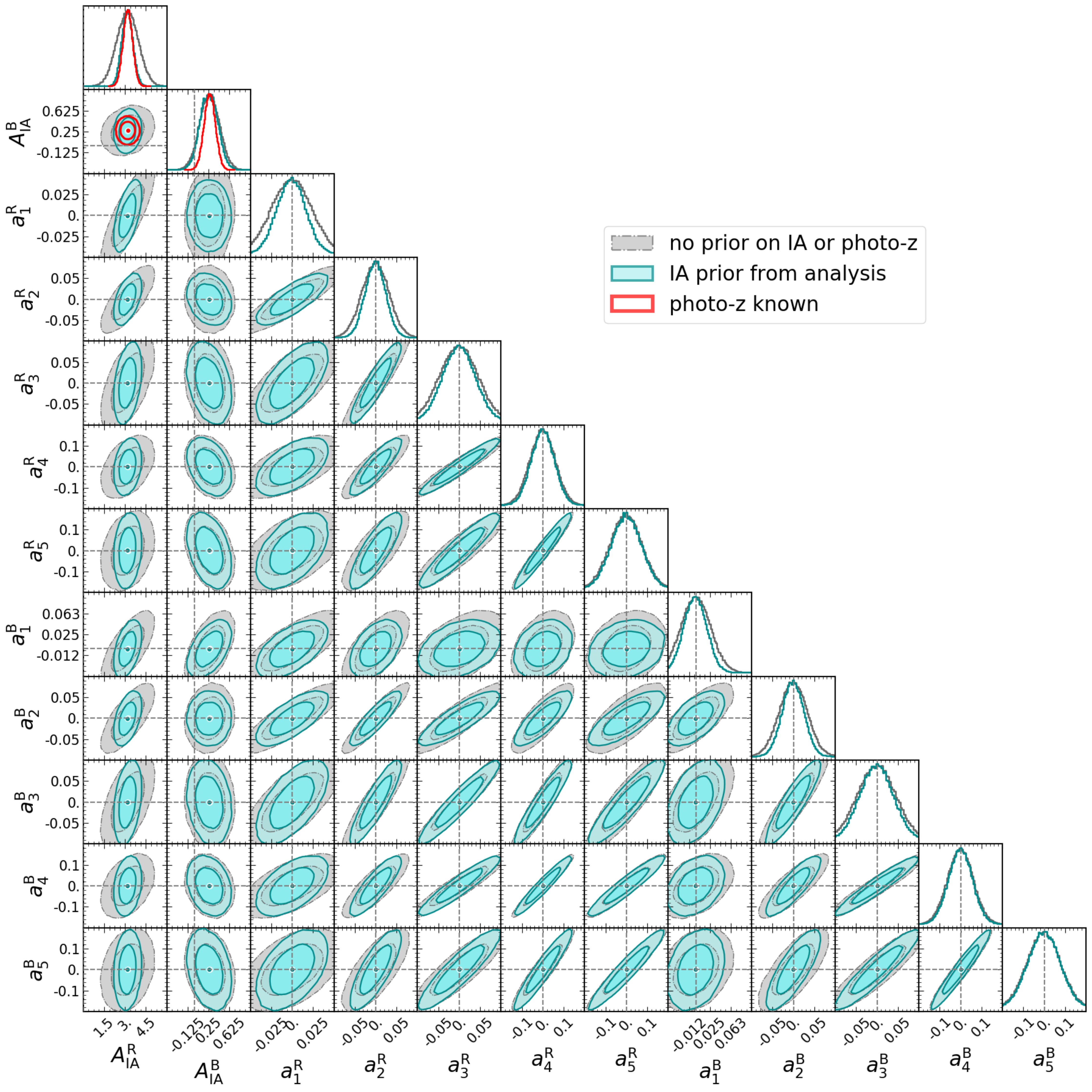}
    \caption{The same as Figure \ref{fig:forecast_nobeta}, for nuisance parameters only.}
    \label{fig:forecast_pz_nobeta}
\end{figure*}

\begin{figure*}
	\includegraphics[width=\textwidth]{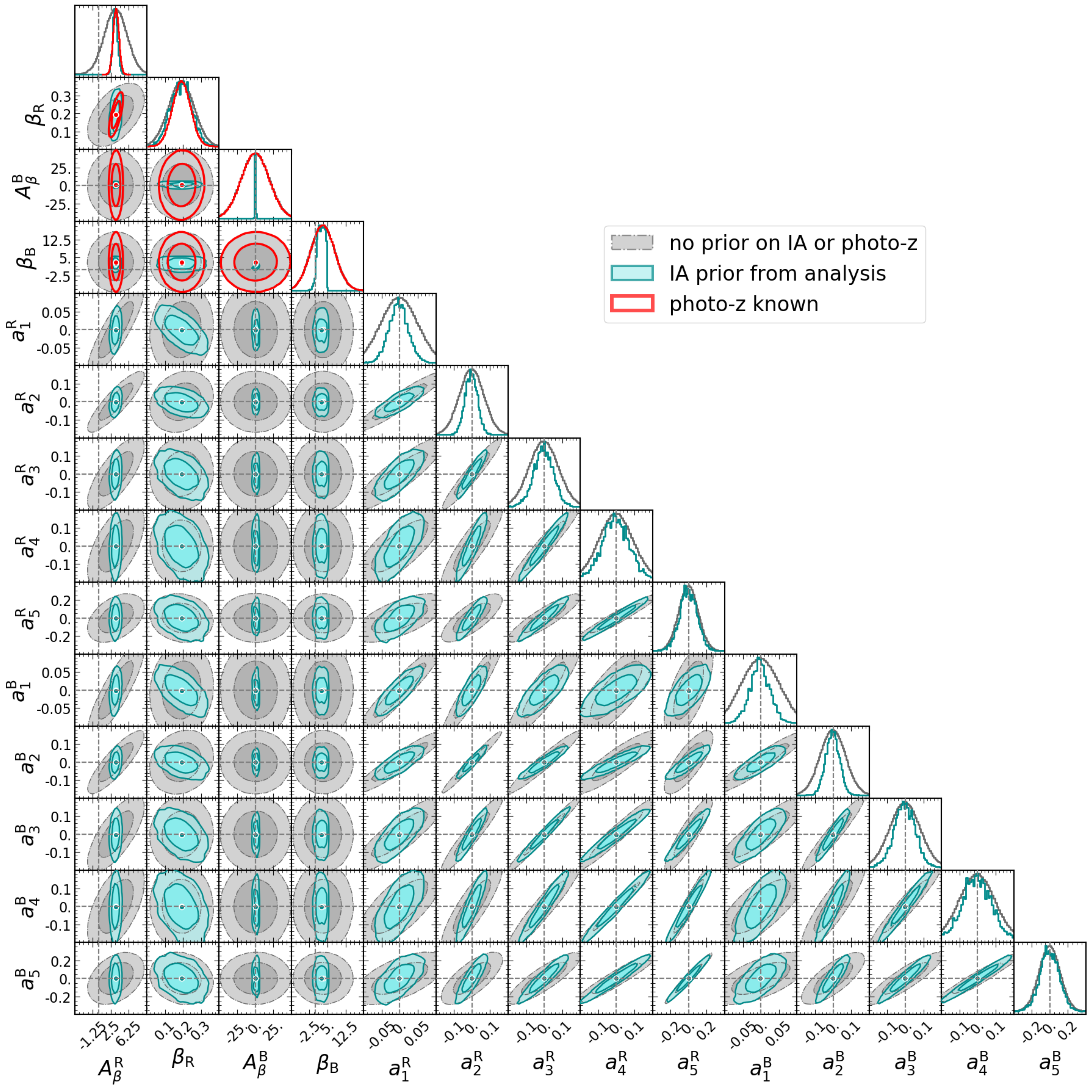}
    \caption{The same as Figure \ref{fig:forecast_beta}, for nuisance parameters only.}
    \label{fig:forecast_pz_beta}
\end{figure*}

Here we include our Fisher forecasted constraints for all IA and photo-$z$ nuisance parameters considered (Section \ref{sec:forecasting}), before and after application of our derived IA model priors (Section \ref{sec:results}). Figures \ref{fig:forecast_pz_nobeta} \& \ref{fig:forecast_pz_beta} accompany the LA/LA-$\beta$ model forecasts of Figures \ref{fig:forecast_nobeta} \& \ref{fig:forecast_beta}, respectively.

\label{lastpage}
\end{document}